\documentclass[12pt, a4paper]{article}

\usepackage[top=2cm, bottom=2cm, left=3cm, right=1.5cm]{geometry}
\usepackage{setspace}
\newcommand{\tablefont}{\fontsize{12pt}{10pt}\selectfont}

\usepackage[utf8]{inputenc}       
\usepackage[T1]{fontenc} 
\usepackage{amsmath}          
\usepackage{amssymb}       
\usepackage{graphicx}   
\usepackage{booktabs}     
\usepackage{caption}
\usepackage{subcaption}
\usepackage{dcolumn}
\usepackage{threeparttable}
\usepackage[titletoc]{appendix}  
\usepackage{tabularx}
\usepackage{makecell}
\usepackage{amsfonts}
\usepackage{amssymb}
\usepackage{graphicx}
\usepackage{rotating}
\usepackage{ragged2e}
\usepackage{array}
\usepackage{pdflscape}
\usepackage{rotating}

\usepackage{longtable}
\usepackage{multirow}
\usepackage{wrapfig}
\usepackage{float}
\usepackage{colortbl}
\usepackage{tabu}
\usepackage{threeparttablex}
\usepackage[normalem]{ulem}
\usepackage{makecell}

\newcommand{\ind}{\perp\!\!\!\!\perp} 

\usepackage{amsthm}
\theoremstyle{definition}
\newtheorem{hypothesis}{Hypothesis}

\usepackage{natbib}
\usepackage{tikz}
\usetikzlibrary{arrows.meta, positioning}
\definecolor{mygray}{gray}{0.85}
\tikzset{
    hidden/.style = {
        draw = black,
        dashed,
        shape = circle,
        fill = mygray,
        minimum size=8mm
    }
}

\usepackage{hyperref}
\hypersetup{
    allcolors = blue,
    colorlinks=true,
    bookmarksopen=true,        
    bookmarksnumbered=true     
}

\author{
    \textit{Vadim Ustyuzhanin}\thanks{HSE University, Center for Stability and Risk Analysis, Research fellow. E-mail: vvustiuzhanin@hse.ru}
}
\title{Parenthood Penalty in Russia: Evidence from Exogenous Variation in Family Size}
\date{} 

\begin{document}

\maketitle
\begin{abstract}
    The present study aimed to improve upon the existing correlational literature on the parenthood penalty in Russia. An instrumental variables approach based on sibling sex composition and multiple births was employed alongside difference-in-differences designs to analyze rich census and longitudinal datasets. To the best of the authors' knowledge, this is the first study to provide causal estimates of the effect of fertility decisions on subsequent labor market outcomes for mothers and fathers in contemporary Russia. The study's primary finding is that, in contrast to the approximately 10 percent long-term motherhood penalty observed in developed countries, the causal impact of childbearing on women's employment in Russia is most significant in the first year after birth, reducing employment by around 15 percent. This penalty then rapidly declines to a modest 3 percent once children reach school age. The analysis indicates an absence of a systematic fatherhood penalty in terms of employment, although a modest increase in labor supply is observed.
\end{abstract}
\textbf{Keywords:} Parenthood penalty; Motherhood penalty; Fatherhood premia; Instrumental variables; Russian labor market.\hfill
\\
\\
\textbf{JEL:} J13, J22 
\thispagestyle{empty}

\newpage
\section*{Introduction}\label{sec:intro}

“A child is for the absolute majority of Russian citizens a one-way ticket to poverty, a second child is a one-way ticket to destitution,” said Michael Delyagin, deputy chairman of the Russian Parliament Committee on Economic Policy \citep{delyagin}. One might continue with this analogy saying a third child is a ticket to the baggage compartment of a train headed for bankruptcy. Indeed, a significant body of research in academic literature has consistently demonstrated that the birth of a child has a substantial negative impact on maternal employment and income levels in different cultural settings \citep[e.g.][]{meta1, meta2}, the phenomenon commonly referred to as the 'motherhood penalty'. Conversely, paternal employment and income levels are sometimes found to increase in response to childbearing \citep{rus_fathers_2020, chn_penalty}. In turn, 'parenthood penalty' has been studied in the russian context too, while not frequently and focusing on 'maternal' one. A small amount of extant research demonstrates an indisputable association between childbirth and a decline in employment and earnings \citep{rus_penalty_2007,rus_penalty_2008,rus_penalty_2017,rus_penalty_2021, rus_penalty_2023}, both in the immediate postpartum period and on an average comparing women with and without children. For instance, \citet{rus_penalty_2007}, who were among the first researchers to analyze Russian data from the early 2000s, found a permanent motherhood penalty of approximately 8-11\% on monthly earnings. Later research support this results: \citet{rus_penalty_2017} showed a 4\% constant reduction in earnings, explaining smaller amount of penalty by recent russian pronatalist policy, while \citet{rus_penalty_2021} found even higher penalty of 17\%. 

Thus, fertility has been demonstrated to exert a detrimental influence on the labor market, resulting in the constant exclusion of a substantial proportion of labor resources from the economy. Moreover, a reduction of employment as well as income of women due to childbearing is a common explanation for gender gap, which not only increases inequality, but also leads to ineffective of women's human capital use in the economy, thereby turning the problem from micro- to macro-level. 

However, while the correlation between childbearing and labor market outcomes appears robust, the causal interpretation of these findings warrants careful consideration. A substantial proportion of extant research in the Russian context is predicated on methodologies that may be susceptible to endogeneity, thereby resulting in an exaggerated negative impact of childbearing. The primary issue is the selection bias. This phenomenon may be intertwined with pre-existing career aspirations, earning potential, or unobserved characteristics that independently influence labor market performance as well as desire for children \citep[eg][]{theory_4}. For instance, women who anticipate or prefer less intensive labor market participation might be more inclined to have more children and vice versa. Consequently, a simple comparison between mothers and childless women, or between mothers with different numbers of children, may overstate the negative impact of children themselves. The second problem pertains to reverse causality. Indeed, researchers in Russia have shown that, on the one hand, the number of children is negatively affected by mother's employment and earnings \citep{revcause_2, revcause_3},
\footnote{Interestingly, the results regarding the effect of employment on fertility in Russia remain controversial. While \citet{revcause_3} found positive effect of employment on the decision to have a child, \citet{revcause_2} presented negative effect using the same data.}
but on the other hand, number of children decreases mother's employment and earnings \citep{revcause_1}. In other words, fertility and economic decisions are co-determined and "since fertility variables cannot be both dependent and exogenous at the same time, it seems unlikely that either sort of regression has a causal interpretation." \citep[p.~451]{angrist_orig}

Furthermore, the direct application of findings from other countries and cultural settings to the Russian context may be misleading for several reasons. Firstly, the presence of a substantial public sector has been demonstrated to engender distinct employment dynamics and disparate levels of job security \citep{pub_sector_sec,pub_sector_sec_2} in comparison with predominantly market-driven developed economies, from which the majority of causal research on fertility and labor market outcomes comes. Secondly, the implementation of active pronatalist policies in recent years may interact with labor market outcomes in specific ways not observed elsewhere \citep{pronatalist_pol,pronatalist_pol_2}. The situation is further complicated by the enduring soviet legacy, which promoted high female labor force participation alongside state-supported childcare \citep{sov_legacy,sov_legacy_2}. However, this legacy also embedded traditional gender roles that continue to influence contemporary family and work dynamics \citep{genderroles_rus,theory_cs_1}. These distinct features demand a focused analysis on Russia in order to comprehend the specific mechanisms and magnitudes of the parenthood penalty since results from other parts of the world might be invalid in russian context. 

The present study aims to address these critical gaps by focusing on the establishment of the causal effect of childbearing on labor market participation in Russia for both fathers and mothers. The research contributes to the field in two ways. From a theoretical point of view, the novelty lies in the utilization of instrumental variables and difference-in-differences designs with census and longitudinal Russian data. From a practical point of view, the study serves to refine existing estimates of the effects of fertility on the labor market outcomes in Russia, thereby providing better estimates of the impact of childbearing on economic behavior of parents.

The rest of the paper is organized as follows. Section \ref{sec:theory} presents universal theoretical explanations of the negative effect of childbearing on labor market outcomes. Consequently, Section \ref{sec:endogeneity} examines prior research about Russia, with a particular focus on endogeneity concerns and potential inconsistencies. In Section \ref{sec:hypotheses}, hypotheses of the study are formulated in accordance with the theoretical background and the identified problems of previous research. Next, Section \ref{sec:methods} details the empirical strategy addressing endogeneity concerns by introduction of instrumental variables and difference-in-difference designs. Subsequently, the choice of data and estimators is described. Section \ref{sec:results} presents the results obtained by 2SLS models and panel matching. Section \ref{sec:discussion} discusses the findings and limitations of the study as well as compares Russia with other developing and developed countries in terms of parenthood penalty. Finally, Section \ref{sec:conclusion} concludes the paper.

\section{Theoretical background}\label{sec:theory}

The negative impact of childbearing on women's employment and earnings is a well-established phenomenon that has been extensively researched from both theoretical and empirical standpoints \citep{meta1,meta2,theory_1,theory_2,theory_3,theory_4,theory_5}. The parenthood penalty can be defined more formally as the systemic disadvantages that parents face in the labor market in comparison to those who do not have children. It is also possible to incorporate additional penalties, encompassing various aspects of well-being, such as health, life satisfaction, and social life; while these components have received  less attention within academic literature. This section is predominantly focused on the economic impact of childbearing, while also addressing other effects. In many respects, the theoretical justifications for the penalties can be divided into two non-contradictory theories, namely rational choice theory and status-based discrimination theory.

The rational choice explanation is based on the classical economic idea that individual's earnings are a function of accumulated skills, education, and work experience, which can be concisely called "human capital". An increase in accumulated human capital is associated with better employment opportunities and higher wages. In turn, childbearing is hypothesized to decrease both women's human capital and further investments to it due to several reasons. Firstly, mothers tend to disrupt their career trajectories with periods of pregnancy and few years after childbearing, that leads to less work experience and training compared with childless women, who choose career \citep{theory_4}. Moreover, because of anticipation of childbearing future mothers tend to invest less in education and training due to perceived lower returns on these investments given potential work interruptions and reduced working time \citep{econ_theory_interrupt}. Additionally, labor-market interruptions can make skills outdated and less effective, thereby decreasing productivity in general, though it should be actual foremost for mothers who already have huge investments to skills -- tertiary education \citep{penalty_by_educ}. Secondly, mothers choose "family-friendly" employment, which is less financially rewarding, but provides greater security and the opportunity for a more balanced 'work-family' trade-off \citep{econ_theory_friendly}. The public sector is often recognized as a such "family-friendly" employer, and mothers are frequently over-represented in it, because of their preferences for security and stability over financial incentives \citep{pub_sector_sec_2,theory_1}\footnote{An excellent example of this phenomenon is provided by an in-depth interview with a russian mother who explains the preference toward public sector employment, as documented in an excellent paper by \citet[p.~202]{soc_theory_status_rus}:  “A steady paycheck... availability of this big vacation, which with children is very convenient ... sick leave, that is, I can go on sick leave at any time, even if the child has a little cough, and completely sit it out, which, I know, is not welcome in commercial organizations”.}. For the same reasons mothers or further mothers might refrain from pursuing promotions that entail extended working hours or increased responsibility\footnote{This may be especially applicable in the context of Russia, where further mothers may prioritize lower-paid roles, whereas employers are bound by a strict yet advantageous labor protection laws designed to support mothers, offering substantial paid leave and guaranteed job security. More on this, as well as examples from russian labor code, see Section \ref{sec:hypotheses}.}. Thirdly, some authors hypothesized that motherhood should lower women's productivity. This is possibly due to mothers redistributing their labor supply and effort from professional activities to housework and childcare, which in turn leads to reduced investment in human capital and less career orientation \citep{theory_3}\footnote{However, children themselves can be considered as economic agents, who would in turn take part of housework, which leads to increased mother's labor supply. Although this explanation may be applicable to traditional societies, in an industrialized world, children are not regarded as workers, but rather as a form of investment and economic good (speaking economically) that is not expected to generate direct economic returns \citep[for a discussion see, for ex.,][]{becker1960}.}. Thus, the motherhood penalty can be interpreted as a consequence of rational choice for relatively more compatible with raising children occupation as well as lower human capital accumulation. 

However, there is a critique of such straight economic explanation that can be called a status-based discrimination explanation. It was shown that mothers might be punished both in terms of wage and employment because of employers' aspirations about lower productivity of mothers compared to childless women \citep{theory_2}. In that case, children play a role of a signal for employers that woman might be less prone toward career and would ask for lower hours, more flexible schedule etc\footnote{To illustrate, I again refer to the interview by \citet[p.~201]{soc_theory_status_rus}: “I went to so many of these interviews... Everyone was happy with my age, education, knowledge of computers. As soon as you announce that you have a child of two or three years old, their gaze immediately dims...”}. Explanations for this phenomenon may be attributed to the fact that childbearing changes the status of women, who are now primarily assigned "social" role and responsibilities for raising and caring for children \citep{soc_stereotypes,soc_theory_status}, as opposed to achieving self-realization through career advancement. In this regard, other potential statuses of women in society (e.g. graduated, young, specific nationalities etc.) may be in the background, while maternity completely shapes the perception of women by others \citep{soc_theory_status}. This results in biased expectations of women in the workplace and underestimation of their abilities and motivation as workers. Moreover, mothers are usually associated with "warmth and caring" \citep{soc_stereotypes} and may be discriminated against when being considered for leadership roles or simply being a responsible worker because of the masculinising attributes of leadership and hard work \citep{soc_theory_status_2}\footnote{It should be noted that it is not only mothers who are discriminated against in this way, but all women as outlined by the role congruity theory developed by \citet{eagly2002role}. However, in line with previous discussion, one might expect mothers to experience more discrimination than childless women.}. Interestingly, even if mothers are proved to be successful and efficient in their positions, they may still encounter discrimination on the basis of their being perceived as "less warm, less likable, and more interpersonally hostile than otherwise similar workers who are not mothers" \citep[p.~616]{soc_theory_status_3}. Consequently, status of "mother" leads to lower employment and earnings, a pattern that should be particularly evident in roles characterized by intensive work schedules and significant responsibility. 

For fathers (and partners of women with children), the literature is more likely to report on either the absence of a penalty or the presence of a "premia" when men with children have a higher level of employment and earnings \citep{theory_fathers_1,theory_fathers_2,theory_fathers_3}. There are several explanations for this phenomenon, some of which are related to theories that posit a penalty for women in relation to childbearing. Firstly, it is argued that the stereotype that men are less efficient and will devote their working time to caring for a child is not valid in contrast to mothers. Conversely, society may associate the social role of the father with greater motivation to work in order to support the family -- wife and children \citep[see][for a comprehensive review on the perceived status/role of fathers]{father_role}\footnote{It is important to note the paucity of research on the potential division between biological fathers and stepfathers (or so-called  residential fathers). Later group, however, constitute a significant share of the population of interest both in the world and in Russia.}. 
Indeed, the "father" role is frequently perceived not as a "caregiver" and "housekeeper", but rather as a "breadwinner" or "provider",  who must work hard and, therefore, will be more productive and responsible than childless man \citep{father_role_2}. Secondly, rational choice theory posits that if the investment of time in child care is necessary, this should be undertaken by the less productive parent, who is usually mother; while father, on the contrary, increases the supply of labor to compensate for the losses of the household.

Thus, status-based discrimination explanation and rational choice theory concur with the same conclusion: women will be employed in more family-oriented but lower-paying occupations, or will be excluded from the work force at all. This effect is quite persistent across different cultures and was shown to work in Russia too \citep{soc_theory_status_rus}, that is further supported by strict russian labor protection laws, which are very supportive of mothers as outlined below in Section \ref{sec:hypotheses}. In turn, both theories also concur with the same expectation about effect for fathers: it should be positive, because fathers are treated as "breadwinners" who should respond to childbearing by increasing labor supply to support the family (both wife and children) and to compensate the decrease of wife's labor supply. 

\section{Previous research and potential problems}\label{sec:endogeneity}

The issue of the paper's opening section, which concerns the extent to which the findings of studies conducted in other countries can be considered valid for Russia, remains unresolved. This is due to the limited number of studies that have been conducted using russian data. However, the extant literature has demonstrated that there is a significant negative effect of fertility on women's employment and income \citep{rus_penalty_2007,rus_penalty_2008,rus_penalty_2017,rus_penalty_2021,rus_penalty_2023}, while the "premium" for fatherhood being either minimal or non-existent \citep{rus_fathers_2020}.

However, previous studies that aimed to identify the constant effect of childbearing – that is, not during the initial years following birth, but until the child attains adulthood – may be subject to severe bias, resulting in an exaggeration of the negative impact due to self-selection issues and reverse causality. The following discussion will address this issue in more detail.

Without loss of generality, the empirical estimand in the previous research is the effect of a childbearing on economic activity of parents -- mothers or fathers. Let $Y_i$ represent the dependent variable (labor force participation or earnings) for individual $i$, and let $d_i$ be a binary treatment variable denoting the fact of childbearing. Then, using the potential outcomes' framework \citep[see, for ex.,][]{rubin_pot}, the individual causal effect $\tau_i$ can be presented as:

\begin{equation}
    \tau_i = Y_i(1) - Y_i(0)
    \label{eq:main1}
\end{equation}
where $Y_i(1)$ and $Y_i(0)$ show the potential outcomes under possible values of treatment $d_i = 1$ (there is a child) and $d_i = 0$ (no child) respectively for the same individual $i$. The overall effect is aggregation of individual effects from \eqref{eq:main1} across the population -- mothers or fathers:

\begin{equation}
    \tau = \mathbb{E}(\tau_i)=\mathbb{E}[Y_i(1) - Y_i(0)]
    \label{eq:main2}
\end{equation}

A usual strategy exploited by previous research is to estimate the $\tau$ via ordinary least-squares (OLS) estimator, assuming the following data-generation process:

\begin{equation}
    Y_i = \beta_0 + \tau D_i + \mathbf{X}_i' \theta + \varepsilon_i
    \label{eq:ols}
\end{equation}
where $Y_i$ is an observed outcome under observed treatment $D_i$ (which is a realization of potential outcome $d_i$) and $\mathbf{X}_i'$ is a vector of some control variables. However, $\hat \tau_{OLS}$ from 
\eqref{eq:ols}\footnote{Hereinafter a hat sign $\hat{}$ denotes the parameter estimate. Thus, $\tau$ is a real true effect, while $\hat \tau$ is its estimate that might differ, meaning at average $\tau - \hat \tau \neq 0$.} 
is consistently estimated if, and only if, $D_i$ is (conditionally) exogenous and, then, as good as (conditionally) randomly assigned across population. In other words, $\hat \tau_{OLS} = \tau$ when $\mathbb{E}(D_i \varepsilon_i) = 0$ and $D_i \ind \varepsilon_i | \mathbf{X}_i$. However, in case of the effect of childbearing this assumption probably does not hold for several reasons. 

Firstly, there can be a "self-selection problem": individuals choose to have children not randomly, but based on a multitude of factors that might be strongly correlated with economic activity. Returning to the potential outcome framework, it means that actual treatment $D_i$ is correlated with potential outcome $Y_i(d)$, which contradicts exchangeability assumption and makes causal identification implausible: $\mathbb{E}(Y_i(d)|D_i = d) \neq \mathbb{E}(Y_i(d')|D_i = d')$. To simplify, in any causal analysis the main problem lies in estimation of counterfactuals. In \eqref{eq:main1} one does observe only one of potential outcomes: for woman $i$ we see, say, income when she has a child (so she is under treatment and observed outcome equals potential one: $Y_i = Y_i(1)$), but there is no possibility to collect data on the exactly same woman $i$ at exactly the same time, when she does not have a child (so when she is under control and her observed outcome equals $Y_i(0)$). If childbearing is randomly assigned across women, we still do not observe second potential outcome, but at average treatment and control groups are similar (they are "twins"), and one can assume that control group represents absent in reality potential outcome $Y(0)$. However, if childbearing is not randomly assigned, observations in control group does not represent the absent potential outcome, because now control group is not a twin, but a distant relative at best. Therefore, a comparison in such a case is meaningless by its very nature, as it fails to address the question of what the outcome would have been in the absence of the child. The usual strategy is to include control variables to reach conditional randomness. However, in case of childbearing is it possible to include sufficient set of covariates to claim random assignment of childbearing? It was shown that mothers dramatically differ from non-mothers by career ambition, innate ability, risk aversion, or preferences for market work versus family responsibilities. For instance, if women with inherently lower earning potential and weaker labor market attachment are more likely to become mothers, a comparison that fail to control for such unobserved factors would erroneously attribute mother's lower earnings primarily to children, when in fact, pre-existing differences play a substantial role.

Secondly, both fertility and economic activity might be co-determined. Indeed, in economic research number of children is a usual independent variable that affects labor outcomes, while in demographic research it is a dependent variable that is affected by same variables \citep{baizan2020linking}. Indeed, for the Russia it was shown that the number of children is negatively affected by mother's employment and earnings \citep{revcause_2, revcause_3}, but on the other hand, number of children decreases mother's employment and earnings \citep{revcause_1}. Therefore, "since fertility variables cannot be both dependent and exogenous at the same time, it seems unlikely that either sort of regression has a causal interpretation" \citep[p.~451]{angrist_orig}. To simplify, the reverse causality leads to self-reinforcing cycle, highly inflating the negative effect of childbearing, which now also includes some part of the negative impact of employment on childbearing. 

\begin{figure}[htbp]
    \begin{minipage}{1\textwidth}
        \centering
            \begin{tikzpicture}[
                node distance=1.5cm and 1cm,
                >=Stealth,
                every node/.style={draw, circle, minimum size=8mm}
            ]
            
            \node (D) at (-1.5,0) {$D_i$};
            \node (Y) at (3.5,0) {$Y_i$};
            \node (X) at (1,2) {$\mathbf{X}_i$};
            \node[hidden] (U) at (1,-2) {$\mathbf{U}_i$};
            
            \draw[->] (X) edge (Y)
                      (X) edge (D);

            \draw[->, blue] (D) edge node[below, draw=none, fill=none] {$\tau$} (Y);
            
            \draw[->, dashed, red] (U) edge (D)
                                   (U) edge (Y);

            \draw[->, dashed, red] (Y) edge [bend right=25] (D);
            
            \end{tikzpicture}
        \caption{DAG for 'classical' model from \eqref{eq:ols2}.}
        \label{fig:dag1}
    \end{minipage}\hfill
    \small{\textit{Note: U denotes unobserved characteristics, X denotes observed (and controlled) characteristics; $\tau$ denotes the effect of interest; arrows show causal paths; dashed lines represent biasing paths.}}
\end{figure}
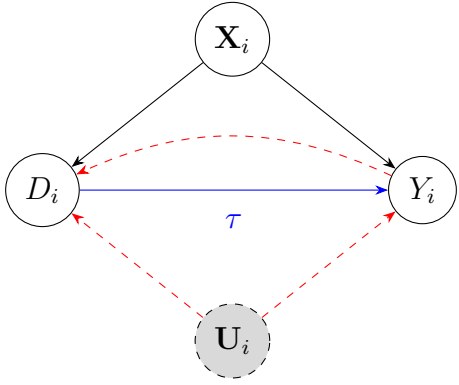

To make the point clearer, consider the following Direct Acyclic Graph (DAG)\footnote{See \citet{DAGs} as an introduction to DAGs for causal identification.} in Figure \ref{fig:dag1} (for simplicity, employment is chosen as a dependent variable). Fertility affects employment that is depicted by arrow, which is under study (it is effect $\tau$). However, there are two main problems. Firstly, there is a vector of unobserved factors $U$ that affects both fertility and employment. This includes (but is not limited to) career ambition and preferences: individuals highly focused on career advancement might choose to have fewer children and to concentrate on professional life, while others might choose the opposite -- to work less and to have more children. Thus, mothers with a relatively large number of children would have worked less anyway and vice versa. Unfortunately, it is almost impossible to control for such a factor, while education may be considered as a proxy for it. Nevertheless, this variable can be not confounder, but a collider that is endogenous and is highly affected by the fertility choice, thereby bringing us to initial problem of omitted unobserved factors. As was shortly summarized by \citet[p.~23]{theory_5}, classical econometric approaches as OLS from \eqref{eq:ols} fail to address these problems "because motherhood has been held to influence both women’s own initial education investments and labor market entry choices and employers’ attitudes and practices with respect not only to recruitment but also to pay and promotion practices and even work organization and job design." Thus, there is no option to consistently estimate the causal effect of fertility $\tau$ since there is a huge set of unobserved covariates. Moreover, there is a problem of reverse causality that is depicted by the arrow from employment to fertility. Since the sign of the arrow is known to be negative as well as the sign of opposite arrow, the $\hat \tau_{OLS}$ should be negatively inflated due to self-reinforcing cycle. 

Hence, using notation from \eqref{eq:ols}, the model under data generation process from Figure \ref{fig:dag1} is:

\begin{equation}
    Y_i = \beta_0 + \tau D_i + \mathbf{X}_i' \theta + (\epsilon_i + \mathbf{U}_i' \beta)
    \label{eq:ols2}
\end{equation}
where $(\epsilon_i + \mathbf{U}_i' \beta)$ is $\varepsilon_i$ from \eqref{eq:ols} and $D_i$ is some function of $Y_i$. Obviously, $Cov(D_i, \varepsilon_i) \neq 0$ since at least $Cov(D_i, \mathbf{U}_i) \neq 0$. Therefore, OLS model fails to produce consistent (and unbiased) estimates of $\tau$. Moreover, from the DAG and discussed literature one can conclude that the bias is negative, so $\hat{\tau} - \tau < 0$ at average, meaning production of highly pessimistic results. 

Based on the discussion above, I suggest that the estimates of motherhood penalty in Russia, ranging from 4\% to 17\%, is an example of such pessimistic results that stem from fail to address reverse causality and self-selection problems. The following discussion will address the findings and methodological pitfalls of each of the previous studies (for a compact summary, see Table \ref{tab:previous_research} below)\footnote{It is essential to emphasize that my intention is not to demonstrate the inherent inferiority of the works under discussion and so forth. The objective is to highlight the limitations of past studies in interpreting results as causal, despite the notable quality of that papers. I therefore ask readers to excuse the somewhat arid tone of the narrative and the failure to emphasize the merits of the papers during the description thereof.}.

The first attempt to quantify the size of the motherhood penalty was made by \citet{rus_penalty_2007}, who exploited Russian Longitudinal Monitoring Survey -- RLMS (2003-2005). The sample is composed of adult women aged between 18 and 49 who have maintained paid employment throughout the observed period and whose level of education is at least secondary. The term 'mother' is then operationalized as a woman who has underage children living with her. Subsequently, the authors transformed the initial data and employed only averages within individuals for the specified period, transitioning from a panel to a cross-sectional setting. The OLS model was finally estimated, with the dependent variable being the wage per month (in current prices) and the independent variable being a categorical variable for the number of children (0, 1, 2, more than 3). In essence, the authors employed the identical model specification as in aforementioned equation \eqref{eq:ols2}, incorporating as control variables mother's age, the age of the youngest child (if applicable), educational level, occupation type, work experience, maternal status, and geographical location (urban/rural). The estimated penalty for one child is 11\%, for two it is 7\% (interestingly, this estimate is insignificant), and for more than three it is approximately 33\% to 40\%, depending on the model specification. In line with previous discussion, these estimates might be considerably biased by self-selection problem, since there is no control for selection procedure. Additionally, occupation type (public/private sector) and education level probably are colliders since they depend on fertility choices, meaning one more source of bias, which inflates negative estimates. Finally, reverse causality is not addressed.

The next study by \citet{rus_penalty_2008} almost replicates the result by \citet{rus_penalty_2007}, but additionally National Survey of Household Welfare and Participation in Social Programs (NOBUS) was used as well as empirical strategy was expanded. Firstly, authors exploited the same research design as before, but then additionally did instrumental variables' estimation, using the presence of a non-working woman over 49 years of age (grandmother) in the household and income of other household members. The final estimate by instrumental variables is about 11\%, while OLS shows about 8\% penalty. Section \ref{sec:iv_theory} describes instrumental variables approach and its assumptions in detail, stating it is indeed capable of addressing the selection bias as well as reverse causality. Nevertheless, in the case of instrumental variables selected by \citet{rus_penalty_2008}, the assumptions are likely to be violated, which leads to even more bias than unadjusted OLS estimation. The instrumental variables are presumed to affect independent variable -- childbearing, -- yet it is further assumed that they do not affect the outcome -- wage. Do income of other households and the presence of grandmother affect mother's behavior on labor market? I state that the answer is positive: in cases where the income of other household members is sufficiently high, it is probable that women will opt for reduced or even non-participation in the workforce; conversely, in households with the presence of a grandmother, it is conceivable that women may choose to increase their labor force participation, as the grandmother may undertake the housework in their stead. Consequently, the selected variables cannot be treated as instruments and their application does not resolve the issue, but rather exacerbates it\footnote{Even if these instruments are exogenous and restriction criteria are satisfied, they appear to be weak: the presence of grandmother has no effect on the number of children and only household income is significant. The utilization of weak instruments has the potential to result in severe bias \citep[e.g.][see also Section \ref{sec:iv_theory} of the presented paper]{weak_inst_1}.}. 

The study by \citet{rus_penalty_2017} was presented almost 10 years after the publication of the two preceding studies. The authors utilised RLMS data from 2014, focusing on employed women within the age range of 20 to 44. In order to address the selection bias, the researchers employed inverse probability weighting (IPW) to achieve a balance between mothers and childless women according to observed characteristics. They then proceeded to apply OLS model, with the dependent variable being the monthly wage. The application of weighting serves to minimize the disparity between the treated (mothers) and control (childless women) groups, thereby ensuring the recreation of a quasi-randomness of childbearing assignment. However, this approach might exclude observed confounding, while it is possible that confounding may be unaltered. The set of covariates includes the following: age, place of residence, marital status, health (as measured by the presence of chronic diseases), and some occupation parameters. Regrettably, these factors alone are inadequate for the control of selection into mothers, as all of them may be regarded as already influenced by underlying women's anticipation, as outlined in the theoretical section above. It therefore follows that while the IPW will reduce bias due to observed imbalances, it will not solve the problem of unobserved imbalance between groups \citep[see][]{king_biases}. Hence, estimates are probably inconsistent due to selection bias. 

A noteworthy study by \citet{rus_penalty_2016} examines 22 industrialized countries, including Russia. Authors utilised RLMS data of 2000/2001 provided by the Cross-National Data Center in Luxembourg, which employs a sample of employed women between the ages of 25 and 45. The research design in this study is analogous to \citet{rus_penalty_2017} and \citet{rus_penalty_2008}, but a Heckman model is utilised to address the selection into motherhood problem as the first stage. The resulting model is for all countries, with country effects being modelled through random slope of the number of children. Additionally, a series of macro-level control variables are incorporated, the majority of which pertain to government policies designed to support motherhood such as the length of job-protected leave due to childbearing, publicly supported infrastructure for childcare etc. In contrast to the extant literature on the subject which focuses on Russia, authors found only an insignificant effect of number of children on wages. Nevertheless, the potential problems of this study are comparable to those of previous studies due to the usage of analogous data and a similar research design, while aggregated cross-cultural model does not allow to specifically study russian case. 

The study by \citet{rus_penalty_2021} is, to the best of my knowledge, the first attempt to examine the motherhood penalty using panel data from the RLMS from 2000 to 2015. In line with previous papers, authors used OLS model, though also introducing individual fixed effects (FE) to control for time-variant factors. The research found that, on average, the wage penalty ranges from 4\% to 17\%, depending on the model specification and dependent variable -- monthly or hourly wages. Whilst the utilization of panel data provides opportunity to control for time-invariant factors using FE, this is not a panacea. The timing of childbirth itself may be endogenous and correlated with transient shocks to an individual's earnings trajectory or employment prospects. This suggests the presence of a set of unobserved factors that changes during the life-course. For instance, a woman might choose to have a child when she perceives her career to be stagnating or if she anticipates a period of lower earnings for other reasons. In such a scenario, the observed decline in post-childbirth earnings might be attributed to these pre-existing negative career dynamics rather than to the impact of motherhood. Conversely, women may schedule births after attaining specific career milestones, a factor which may also complicate the identification of causal inferences. Thus, the presence of unobserved time-varying factors, including changes in motivation, health status, life-satisfaction etc. that differentially affect women around the transition to motherhood, might also lead to inconsistent estimates even within a FE framework. Therefore, while FE models are able to account for persistent individual differences, they may not fully capture the dynamic selection processes or the influence of time-variant unobservable factors that have the potential to confound the estimated impact of motherhood. To sum up, it is hard to argue that "mothers are born, not made", thus illustrating that the usage of FE does not resolve the issue. Moreover, the usage of FE OLS in panel data settings implicitly assumes that there past outcomes (labor market outcome) do not directly affect current outcome; past outcomes do not directly affect current treatment (the presence of child); and past treatments do not directly affect current outcome \citep{imai_fe_causal}. In case of childbearing, all that assumptions are likely to be violated, which leads to biased estimates\footnote{Also, recently \citet{twfe_hetero} highlight one more problem with FE OLS for causal inference in panel data setting. In that case, the ATE is the weighted sum of individual's ATE. If there is treatment effect heterogeneity (i.e. effect of childbearing is not constant for units, years, etc.), FE OLS may reverse the sign of the true effect due to "negative weighting".}. 

Additionally, in aforementioned studies \citep{rus_penalty_2007,rus_penalty_2008,rus_penalty_2017,rus_penalty_2021} there are other potential problems. The first is the inclusion of colliders -- such control variables that are endogenous themselves and affected by fertility and/or labor market outcome. The examples are education and occupation status that are at least partially affected by childbearing (and unobserved anticipated childbearing). Their inclusion might bias the effect of childbearing, since their effect in the model will be partially determined by children. The next already discussed problem is the reverse causality that is not addressed in the previous research and might inflate the negative effect of childbearing. 

The research by \citet{rus_penalty_2023} differs dramatically from the others, exploiting event-analysis with RLMS data from 1994 to 2018. Additionally, authors used not only wage, but also employment as well as monthly worked hours as dependent variables, expanding the analysis of parenthood penalty\footnote{However, authors claimed to find no effect of fatherhood penalty, and throughout the paper they analyzed only motherhood penalty.}. The sample included both mothers and fathers. The paper uses an event-analysis research design, which is an OLS model with FE, as well as dummies for the periods before and after the treatment (childbearing), to recreate the difference-in-difference design (more on this design, see Section \ref{sec:diff_in_diff}), which might effectively address selection bias and produce average effect on treated (ATT) effect. However, the difference-in-difference design (and therefore event analysis) assumes that there are parallel trends between the control and treated groups before treatment adoption. In other words, women who will become mothers some time in the future is a treated group, while women who have the same life trajectories as in treated group (in education, employment, partnership etc.) is a control group. Because of that similarity in pre-treatment period (childbearing), control group is a good proxy how the treated group would have changed over time without treatment. In the case of OLS with period dummies, some support for the PTA might be shown if the dummies for the pre-treatment periods (which show the difference between the treated and control groups in the dependent variable) are insignificant. Unfortunately, this is not the case in the study by \citet{rus_penalty_2023}, where in the majority of the models the pre-treatment dummies are negative and significant (see Tables A3-A9). This might indicate the imbalance between control and treated group and selection bias\footnote{While authors did not discuss that issue, the significant coefficients of $t-1$ dummy might be the product of pregnancy since $t=0$ is a birth. However, in that case there is no possibility to assess PTA, because $t-1$ is a part of treatment, $t-2$ is a reference category, and only $t-3$ remains that is not sufficient to assess PTA. Also, it should be noted that even if there were a few more pre-treatment periods with insignificant dummies, it is just weak support for PTA in case of FE OLS \citep[see][]{twfe_critique_2}. In turn, there are models where $t-1$ dummy is insignificant that, on the one hand, might show the zero effect of pregnancy on worked hours and wage, but on the other may indicate the lack of statistical power of the tests. Interestingly, in the models where $t-1$ dummy is insignificant, all other effects ($t+1$, $t+2$, etc.) are also insignificant, meaning no parenthood penalty. Thus, the problem with statistical power is highly possible.}. 
Additionally, event study regression does not differ in its assumptions from the usual FE OLS used in \citet{rus_penalty_2021}, meaning all other issues remain unaddressed. However, this study provides the most reliable estimates of the motherhood penalty in terms of the number of dependent variables and the research design. Nevertheless, it only focuses on the period immediately after childbirth — five years. Therefore, results cannot demonstrate the long-term effects of childbearing. Given that other authors claim the main negative effects occur immediately after birth, the results by \citet{rus_penalty_2023} cannot be used to assess the long-term impact of parenthood. 

Research by \citet{rus_fathers_2020} is a rare (and the only one to my best knowledge\footnote{However, there is one more research by \citet{rus_fathers_2010} that merits a footnote. It focused on fatherhood premia with usage panel micro data from 2004 to 2007. Whilst author efficiently summarized the "selection-into-fathers" problem and even exploited difference-in-difference design, it is unfortunate that it is impossible to identify the treated and control groups from the text, since the author did not describe them. In the absence of the aforementioned elements, as well as no diagnostics for PTA, any discussion of this research is unfeasible.}) example of the study that focuses on fatherhood penalty in Russia. In the study RLMS data for 2000-2018 period is used, from which only residential fathers (who live with children) in partnership aged 25-59 were selected. Exploiting FE OLS approach as in \citep{rus_penalty_2021}, author found small, but positive effect on earnings for fathers who have one child under the age of 3. Since potential issues with the FE OLS estimator for estimating the parenthood penalty have already been discussed, I do not elaborate the point with selection bias concerns, though author effectively addressed possible "anticipation" problems that violates PTA as well as did not include colliders into the model.

\addtolength{\tabcolsep}{+3pt}  
\begin{sidewaystable}
    \caption{Summary of previous research findings and identification of potential problems.}
    \centering
    \tablefont
    \label{tab:previous_research}
    \begin{tabular}{p{3cm} p{4cm} p{3cm} p{2cm} p{1.5cm} p{6cm}}
    \toprule
    \multicolumn{1}{c}{Reference} &
    \multicolumn{1}{c}{Data \& sample} &
    \multicolumn{1}{c}{Estimator/correction} &
    \multicolumn{1}{c}{Dep. var} &
    \multicolumn{1}{c}{Effect} &
    \multicolumn{1}{c}{Possible problems}
    \\
    \midrule
    
    \RaggedRight \citet{rus_penalty_2007} &
    \RaggedRight RLMS, 2003-2005 (treated as CS); Employed women aged 18-49 &
    \RaggedRight OLS / -- &
    \RaggedRight Monthly wage, ln &
    \multicolumn{1}{c}{-0.08} &
    \RaggedRight Selection bias; Inclusion of colliders: education, occupation status; Reverse causality
    \\

    &&&&&\\

    \RaggedRight \citet{rus_penalty_2008} &
    \RaggedRight RLMS, 2003-2005 (treated as CS) + NOBUS 2003; Employed women aged 18-49 &
    \RaggedRight Switching OLS / instrumental variables &
    \RaggedRight Monthly wage, ln &
    \multicolumn{1}{c}{-0.113} &
    \RaggedRight Weak instruments; Violation of exogeneity and exclusion restriction; Inclusion of colliders: education, occupation status
    \\

    &&&&&\\

    \RaggedRight \citet{rus_penalty_2016} &
    \RaggedRight RLMS, 2000/2001 (CS); Employed women aged 25-45 (russian data as part of cross-cultural dataset) &
    \RaggedRight RE OLS / Heckman model &
    \RaggedRight Monthly wage, ln &
    \multicolumn{1}{c}{Insignificant effect} &
    \RaggedRight Selection bias due to unobserved factors in Heckman model; Inclusion of colliders: education; Reverse causality
    \\

    &&&&&\\

    \RaggedRight \citet{rus_penalty_2017} &
    \RaggedRight RLMS, 2014 (CS); Employed women aged 20-44 &
    \RaggedRight OLS / IPW &
    \RaggedRight Monthly wage, ln &
    \multicolumn{1}{c}{-0.039} &
    \RaggedRight Selection bias due to unobserved factors in IPW model; Inclusion of colliders: education, occupation status, health; Reverse causality
    \\

    &&&&&\\

    \RaggedRight \citet{rus_penalty_2021} &
    \RaggedRight RLMS, 2000-2015 (panel data); Employed women aged 18-45 &
    \RaggedRight OLS / FE &
    \RaggedRight Monthly / hourly wage, ln &
    \multicolumn{1}{c}{-0.104/-0.039} &
    \RaggedRight FE OLS problems for difference-in-difference strategy as outlined in the text; Inclusion of colliders: education, occupation status; Reverse causality 
    \\

    &&&&&\\

    \RaggedRight \citet{rus_penalty_2023} &
    \RaggedRight RLMS, 1994-2018 (panel data); Any women &
    \RaggedRight OLS / Event analysis design  &
    \RaggedRight Employment; monthly wage, ln; hours worked &
    \multicolumn{1}{c}{--} &
    \RaggedRight Violation of PTA
    \\

    &&&&&\\

    \RaggedRight \citet{rus_fathers_2020} &
    \RaggedRight RLMS, 2000-2018 (panel data); Residential fathers in partnership aged 25-59 &
    \RaggedRight OLS / FE and anticipation tests  &
    \RaggedRight Monthly wage, ln &
    \multicolumn{1}{c}{0.03} &
    \RaggedRight FE OLS problems for difference-in-difference design as outlined in the text
    \\

    \bottomrule 
    \end{tabular}
    \begin{tablenotes}[flushleft]
    \small{\item \textit{Note: OLS -- Ordinary Least Squares; CS -- cross-sectional data; FE -- fixed-effects; RE -- random effects; IPW -- inverse probability weighting; PTA -- parallel trends assumption. For a detailed discussion of each of the papers, see Section \ref{sec:endogeneity}.}}
    \end{tablenotes}
\end{sidewaystable}
\addtolength{\tabcolsep}{-3pt}

\section{Hypotheses}\label{sec:hypotheses}

The preceding Sections provided a review of extant published studies on the problem of quantifying the parenthood penalty in Russia, focusing on maternal penalty. The majority of these studies attempted to identify a permanent (or at least long-term) effect of childbirth on the wages of mothers in comparison to non-mothers. Almost all papers, with the exception of one \citep{rus_penalty_2016}, have identified a negative relationship between motherhood and wages \citep{rus_penalty_2007,rus_penalty_2008,rus_penalty_2017,rus_penalty_2021}, with the magnitude of this effect ranging from 4\% to 17\%. This is a considerable effect, given that such an impact is assumed to be constant throughout the greater part of a woman's professional life. Nevertheless, the authors note that this is largely explained by a considerable decline in the first years following childbirth, although a negative effect of motherhood is also observed later in life. Furthermore, the authors demonstrated that the effect can be heterogeneous by educational status, with women who have received a higher education being more susceptible to the penalty \citep{rus_penalty_2017}. This finding is consistent with the results obtained by researchers in other countries \citep{angrist_orig,penalty_by_educ}. However, it should be noted that each of the studies may be subject to a selection problem that is quite difficult to address using classical econometric approaches exploited by previous studies. Meanwhile, an exception is study by \citet{rus_penalty_2023}, where event analysis is employed. However, it is also not without limitations and concerns about violation of the main identifying assumption -- parallel trends. One way or another, it only focuses on the period immediately after childbirth, so results cannot demonstrate the long-term effects of childbearing in contrast to aforementioned. Furthermore, the extant literature has exclusively focused on the wages of employed women\footnote{Studies by \citet{rus_penalty_2023} and \citet{rus_child_ls}  are exceptions, while, one more time, they show only the immediate effect of childbearing.}, thereby excluding a subset of mothers who did not re-enter the workforce following childbirth, and also excluding the unemployed from the control group. Consequently, it is challenging to ascertain the impact of motherhood on labor supply in principle. Specifically, it is uncertain whether the birth of a child leads to a reduction in the labor supply of women and fathers. The theoretical arguments previously outlined provide compelling evidence that childbirth has a significant impact on women's employment, both in the immediate period following childbirth and over the course of their lifetimes. In comparison with other developed countries, however, I expect that the magnitude of employment penalty to be less pronounced in Russia due to the distinct characteristics of its labor market. These include high labor mobility and "downward" flexibility, which result in a relatively low rate of firings during periods recession, because of the prevalence of wage and hours reduction measures as well as large public sector \citep{rus_lab_market_1, rus_lab_market_2, rus_lab_market_3}. However, this expectation is too complicated to be tested there, since the paper focuses on identifying the causal impact of childbearing in Russia, but it will be addressed in the discussion section. In turn, in line with other research, I do not expect any effect on father's employment. These are the core hypotheses of the study that tests the causal effect of the childbearing: 

\begin{hypothesis}
    \emph{Childbearing has a negative effect on maternal employment.}    
\end{hypothesis}

\begin{hypothesis}
    \emph{Childbearing has no effect on employment of mother's husband.}    
\end{hypothesis}

In addition to the primary hypotheses, I expect the effect to be heterogeneous by level of education. However, while other authors have noted that the negative effect of having a child should be predominantly for women with higher education, I am more inclined to believe that in Russia, on the contrary, the effect will be more negative for women without higher education, that is, who are more likely to be employed in less skilled sectors. This can be attributed to a several underlying factors. Firstly, women who have received a higher education are more likely to be employed within the public sector, which is a "family-friendly" employer and will not require mothers to resign in order to avoid the relatively stringent requirements of Russian legislation on paid leave. Whilst russian labor laws are formally strict and protecting, their enforcement is low, especially in private sector \citep{rus_lab_market_3}. This situation is especially pronounced in case of maternity leave and other benefits for mothers (and some for single fathers) in labor market. To illustrate, pregnant woman, mother with child under age of 3 as well as single mother with child under age of 16 cannot be fired (Articles 261, 298)\footnote{All information here and below is sourced from the latest edition (07.04.2025) of the Russian labor code with the indicated particular articles.}; the employer is obliged to reduce the production rate of a pregnant woman while maintaining her salary, and in the case of "unfavorable factors", completely release her from work while maintaining her salary (Article 254); Additional feeding breaks for the child(ren) at least every 3 hours for at least 30 minutes each for women with child(ren) under age of 1.5 (Article 258); lower job responsibilities upon application with save of previous salary for women with child(ren) under age of 1.5 (Article 254); Additional annual leave "at a convenient time" of up to 14 calendar days upon request for mothers with at least 2 children under age of 14; "maternity leave" (Article 255) and "child care leave" (Article 256) are in total of 140 days (these leaves follow one another); after their end woman can immediately go on paid "parental leave" to care for her children until they are three years old, which means that total length of leave is about 1175 days (note, employer must retain mother's position and not fire her during the whole time). 

Furthermore, women who have received tertiary education are more likely to be able to defend their rights before their employers and not face illegal schemes to terminate the employment of women with children, in circumvention of the law. Finally, studies discussed in Section \ref{sec:theory} provided compelling evidence to support the claim that employers with intensive labor usage are more likely to exhibit bias against hiring mothers, and indeed, to dismiss them. This bias should be higher in unskilled roles characterized by strict working schedules. Because of that, I expect women without a tertiary education will face a greater penalty for motherhood compared to women who have received a tertiary education in terms of employment:

\begin{hypothesis}
    \emph{Negative effect of childbearing is more pronounced among mothers without tertiary education.}    
\end{hypothesis}

The hypotheses described above relate to the permanent effect of parenthood. Nevertheless, this research fails to identify the effect of childbearing on wages, which have been a focal point of preceding studies, because of the data limitations (see Section \ref{sec:data}). In defense, it seems to be the only opportunity to examine the causal effect of motherhood on women's economic status, and to move away from correlations that almost certainly greatly overestimate the negative effect of having a child as outlined in Section \ref{sec:endogeneity}.

However, the present study also considers the effect of childbearing on employment and wages immediately after the birth to examine the short- and medium-term parenthood penalty. Such an analysis requires stricter assumptions which makes it less possible to interpret the results as causal in comparison with hypotheses 1-3. In this case, the effect of childbearing on employment and earnings of mothers is hypothesized to be negative, with this effect tending to diminish as the child grows older. This is primarily due to the expiration of maternity leave and the child reaching the age of enrollment in kindergarten and school, at which point the mother is able to allocate more time to her professional commitments. In turn, fathers are not expected to experience lower employment and wages. Thus, the proposed hypotheses are longitudinal analogues of hypotheses 1-2 and are not distinguished as separate propositions. Correspondingly, results derived from alternative data and research designs should be regarded as complementary, with the potential to present a more nuanced perspective. 

Furthermore, it is anticipated that job satisfaction declines significantly among employed women who continue to work after childbearing. This could be primarily attributed to the fact that such women were not eligible to receive the legally established bonuses for childbirth. Concurrently, they are compelled to significantly reduce their leisure time in order to allocate more time to their professional and familial responsibilities. Interestingly, this effect of motherhood has not been examined in previous studies. Finally, it is expected that aforementioned effects will depend on birth order. The present paper puts forward a simple assumption that the first-time motherhood experience will be associated with the most negative consequences in comparison to those who have had no children. In turn, when comparing women with one child and two children, the effect of having a second child on employment and earnings is smaller for two reasons. Firstly, women adapt to their new circumstances, and secondly, many of the negative effects stem from the mother's status, which is imposed immediately after the birth of the first child. However, the birth of a second and subsequent children will no longer have the same strong negative effects, comparing to mothers with one of two children respectively.

\section{Data \& Methods}\label{sec:methods}

In this section the empirical design of research is presented. Firstly, the general analytical strategy is described (Section \ref{sec:emp_strategy}), focusing on the two approaches -- instrumental variables and difference-in-difference. Secondly, Section \ref{sec:estimators} describes the technical implementation of each method and operationalizations. Finally, in Section \ref{sec:data} data sources are presented, namely random samples from russian censuses of 2002 and 2010, which are used for IV analysis, and Russian Longitudinal Monitoring Survey, which is used for difference-in-difference estimates. 

\subsection{Empirical Strategy}\label{sec:emp_strategy}

Next, two potential research designs addressing the problems of regression models discussed in Section \ref{sec:endogeneity} are discussed. Namely, instrumental variables and difference-in-difference approaches, each with their own set of advantages and disadvantages.

\subsubsection{Instrumental variables}\label{sec:iv_theory}

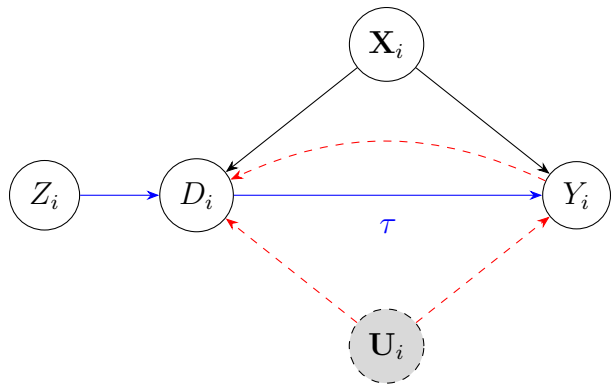
\begin{figure}[htbp]
    \begin{minipage}{1\textwidth}
        \centering
            \begin{tikzpicture}[
                node distance=1.5cm and 1cm,
                >=Stealth,
                every node/.style={draw, circle, minimum size=8mm}
            ]
            
            \node (Z) at (-3.5,0) {$Z_i$};
            \node (D) at (-1.5,0) {$D_i$};
            \node (Y) at (3.5,0) {$Y_i$};
            \node (X) at (1,2) {$\mathbf{X}_i$};
            \node[hidden] (U) at (1,-2) {$\mathbf{U}_i$};
            
            \draw[->] (X) edge (Y)
                      (X) edge (D);

            \draw[->, blue] (D) edge node[below, draw=none, fill=none] {$\tau$} (Y)
                            (Z) edge (D);

            \draw[->, dashed, red] (U) edge (D)
                                   (U) edge (Y);

            \draw[->, dashed, red] (Y) edge [bend right=25] (D);
            
            \end{tikzpicture}
        \caption{DAG for instrumental variables approach.}
        \label{fig:dag2}
    \end{minipage}\hfill
    \small{\textit{Note: U denotes unobserved characteristics, X denotes observed (and controlled) characteristics; $\tau$ denotes the effect of interest; Z is an instrumental variable; arrows show causal paths; dashed lines represent biasing paths.}}
\end{figure}

A possible empirical strategy that gives a chance to consistently estimate the effect of fertility on economic activity that is not suffering from omitted variables problem as well as reverse causality is an instrumental variables approach.  Consider the new version of DAG from Figure \ref{fig:dag1} in Figure \ref{fig:dag2}, where a variable $Z$ is added.  It has a causal effect on fertility, while does not depend on other factors in the model and affects employment only via fertility. Such a variable is called an instrument. In other words, it is a source of exogenous variance in the fertility that does not depend on $\mathbf{U}$ in \eqref{eq:ols2} and enables to estimate the causal effect $\tau$ under some regular assumptions, which are discussed further.

As candidates for instruments for fertility, two variables are used that were already exploited by other authors. The first is the sibling sex mix in families of the fist two children in families with at least two children. This instrument was originally proposed by \citet{angrist_orig}. Underlying idea is straightforward: parents with two children have a slightly higher probability to have a third child if the sex of the first two children are same-sex. For instance, if a couple have two boys (girls), the probability to have one more child is higher than if a couple have boy and girl. This finding is quite persistent across different countries and years, suggesting it works for Russia too: \citet{adsera2011babies} found its significant effect using sample of 13 EU countries from 1994 to 2000; \citet{cruces2007fertility} found the same significant results for Argentina and Mexico, using censuses' data, while \citet{revcause_3} found positive association for Russia using RLMS data. Since the sex of a child is independent of any mothers characteristics (no arrows from $U$ to $Z$ in Figure \ref{fig:dag2}) and, then, randomly assigned\footnote{I do not consider extreme cases, where sex of a child might be a function of some parents characteristics, while it might be plausible in some circumstances. For example, in China during one-child policy the sex ratio rose from classical 105 boys per 100 girls to 150 boys per 100 girls in some regions due to sex-selective abortions \citep[e.g.][]{one_child_policy}. In case of Russia such a scenario is unrealistic, and the sex ratio at birth is stable.}, the exogeneity assumption is hold. The next assumption is exclusion restriction -- the instrument should affect dependent variable only through independent (no arrows from $Z$ to employment). In the case of same-sex instrument, it is highly probable since there is no evidence to expect the effect of children's sex on economic activity of parents not through fertility itself, but by other channels. 

The second candidate for instrument is multiple births, which was first used as a source of exogenous variance by \citet{twins_inst}, while they used its specific version -- twins occurrence. Since the birth of one or two children at once is completely random and does not dependent on any mothers characteristics, it is exogenous\footnote{However, it is known that multiple births are more likely for older women \citep[e.g.][]{twins_and_age} which can be controlled by inclusion of age in the model, thus not violating the assumption of (conditional) exogeneity of the instrument.}. The exclusion restriction also holds, as the birth of an additional child cannot affect economic activity through channels other than fertility.

One more assumption about the instrumental variables is relevance -- $Cov(Z_i, D_i) \neq 0$. More practically, the association between an instrument and variable of interest should be strong enough, because weak link between them may lead to substantial bias in the estimated effect, which does not depend on the sample size. More specifically, the IV estimate via 2SLS model $\hat{\tau}_{2SLS}$ (see Section \ref{sec:2sls} for the specification) converges to 
$$plim \ \hat{\tau}_{2SLS} = 
\tau + \frac{Corr(Z, \varepsilon)}{Corr(Z, D)} \frac{\sigma_{\varepsilon}}{\sigma_{D}} = 
\tau + \frac{Cov(Z, \varepsilon)}{Cov(Z, D)}$$
where $\varepsilon$ is error term, $\sigma_{\varepsilon}$ and $\sigma_{D}$ are the standard deviations of error and independent variable respectively. The fraction is equivalent to the (asymptotic) bias, where the numerator is the bias due to violation of exogeneity and restriction criteria, while the denominator is the strength of the instrument. Therefore, even "small" violations of theoretical assumptions resulting in negligible correlation between the instrument and the error term (i.e. endogeneity) may be dramatically exacerbated by the weak instruments. Fortunately, this assumption is distinct from the two aforementioned assumptions in that it can be empirically tested. Looking ahead, both instruments appear to be quite strong (as in previous research).

The last, but not least, clarification that should be done is that estimation via instrumental variables can be interpreted not as average treatment effect (ATE), but as local average treatment effect (LATE). Given that there are no "defiers" in the sample (monotonicity assumption), who choose not to have children ($D_i = 0$) because of instrument ($Z_i = 1$)\footnote{More formally, $D^{Z=1}_i - D^{Z=0}_i \geq 0 \ \forall \ i$, which means that population is affected by the instrument in the same way \citep[see][]{late1}. Thus, LATE shows not the $\tau$ from \eqref{eq:main2}, which is ATE, but $\mathbb{E}[Y_i(1)-Y_i(0)|D^{Z=1}_i - D^{Z=0}_i \geq 0]$.}; 
instruments are (conditionally) exogenous and are strong enough; and exclusion restriction holds, the estimated $\hat \tau_{IV}$ shows the effect of childbearing on those whose treatment status $D_i$ can be changed by the instrument $Z_i$, so for "compliers" subpopulation \citep[more on LATE, see][]{late1, late2}\footnote{The share of "compliers" in the population for the siblings' sex composition and the multiple births instruments under assumption of no "defiers" are presented below in Section \ref{sec:results}, Table \ref{tab:wald_est}.}. However, LATE is equal to ATE if the effect of childbearing is homogenous across population. I see this as a too strong and unrealistic assumption, because fertility effect likely varies based on many observed and unobserved factors, such as level of education, age, health, income levels, marital status, husband's earnings, mother's anticipations etc.

\subsubsection{Difference-in-difference}\label{sec:diff_in_diff}

In order to enhance the analysis with greater nuance and to examine the effect over time, as well as to undertake a robustness check on the results using a different set and type of data, I utilise longitudinal data, where individuals are tracked over years. Nevertheless, the implementation of IV analysis in such type of data is unrealistic, taking into account aforementioned instruments. The main reason is that such analysis, controlling for unobserved time-variant individual heterogeneity via fixed effects, requires instruments with substantial temporal within variation. In other words, instruments should change from year to year for each individual, while described instruments -- sibling sex composition as well as multiple births -- are time-invariant. Therefore, such variables are incompatible with fixed effects estimators, which absorb all time-invariant variation. However, applying 2SLS model in such setting is still possible using pooled model (the same as \eqref{eq:ols2}, but where $\varepsilon_i$ is now a sum of white noise $\epsilon_i$, unobserved factors $\mathbf{U}_i$ and new component -- individual fixed effect $\alpha_i$) and assuming strict exogeneity, which means independence of observations: so, individuals characteristics and decisions in time $t+1$ does not depend on them in time $t$, $t-1$ and so on. Such an assumption seems highly unrealistic and pooled 2SLS estimator will lead to substantial bias, so another type of research strategy to deal with longitudinal data is exploited.

Difference-in-difference design's main assumption is the parallel trends between control and treated group before treatment. Putting differently, women who will become mothers some time in the future is a treated group, while women who have the same life trajectories as in treated group (in education, employment, partnership etc.) is a control group. Because of that similarity in pre-treatment period (childbearing), control group is a good proxy how the treated group would have changed over time without treatment. Thus, control groups trend is the missing potential outcome from \eqref{eq:main1} that makes possible to calculate the $\tau$ -- $\tau_{DiD}$. In that case $\tau_{DiD}$ is the average effect on treated (ATT), showing:

\begin{equation}
    \begin{aligned}
        \tau_{ATT} = 
        [\mathbb{E}(Y_{1i} | D_{i} = 1) - \mathbb{E}(Y_{0i} | D_{i} = 1)] - \\
        [\mathbb{E}(Y_{1i} | D_{i} = 0) - \mathbb{E}(Y_{0i} | D_{i} = 0)]
    \end{aligned}
    \label{eq:didatt}
\end{equation}
where additional subscript for $Y_i$ shows the time relative to treatment: $0$ - before treatment and $1$ - after treatment. In other words, $\tau_{DiD}$ is the difference between trends in treated and control group. In \eqref{eq:didatt} the simplest version with two periods is demonstrated, but it can be easily generalized to more periods.

As was mentioned before, the main assumption is the parallel trends (PTA) between groups, which also implicitly suggests the absence of time-varying confounding. In case of childbearing and labor market outcomes there is a plethora of unobserved factors that might violate it, if one just finds women for control group who have the same employment history as women in the treated group. Thus, instead of usual PTA, conditional PTA is introduced. In terms of potential outcomes, it means 
$\mathbb{E}[Y(0)_{1i} - Y(0)_{0i} | D_i = 1, \mathbf{X}_i] = \mathbb{E}[Y(0)_{1i} - Y(0)_{0i} | D_i = 0, \mathbf{X}_i]$. 
Thus, the potential trend in case of no treatment is the same in both groups, conditional on covariates. If the set of variables in $\mathbf{X}_i$ is sufficient to construct such groups that there would be no difference in trends before treatment adoption, so $\tau_{ATT}$ from \eqref{eq:didatt} is $0$ before childbearing, then $\hat{\tau}_{DiD}$ is a consistent estimate of $\tau_{ATT}$.

However, such a design requires, firstly, the extensive set of variables $\mathbf{X}_i$ to assume conditional PTA. Secondly, individuals within the control and treated groups should be observed over a long period both prior to and following the adoption of treatment in order to verify similar life-trajectories. Thirdly, it is assumed that there is no time-varying confounding, i.e. individuals are randomly assigned to the control and treated groups and would behave similarly in the scenario of no childbearing. This final assumption is the strongest one, and refers to the self-selection problem discussed above. Whilst the instrumental variables approach is reliant upon clear and partially testable assumptions in order to solve the problem, the DiD design relies upon theoretical arguments that the observed parallel trend over a certain period of time (say, 4 years before the childbearing) is sufficient to eliminate the effect of selection bias.

\subsection{Estimators}\label{sec:estimators}

\subsubsection{2SLS model}\label{sec:2sls}

The usual strategy to estimate the effect of endogenous variable of interest $D_i$ using instrument $Z_i$ is to run Two-Stage Least Squares (2SLS) model as well as provide Wald-estimates. Since both $D_i$ and $Z_i$ are binary variables, the effect $\tau$ can be estimated via Wald-estimator as:

\begin{equation}
    \tau = \frac{\mathbb{E}[Y_i | Z_i = 1] - \mathbb{E}[Y_i | Z_i = 0]}
    {\mathbb{E}[D_i | Z_i = 1] - \mathbb{E}[D_i | Z_i = 0]}
    \label{eq:ivwald_theory}
\end{equation}
which sample analog estimate is $\hat \tau_{Wald}$, where $\hat{\mathbb{E}}[.]$ consistently estimated by sample mean across sub-samples of $D$ and $Z$. It should be noted that while $\hat \tau_{Wald}$ is an unbiased estimate of $\tau$, its variance is too large in case of not strong enough instruments which complicates the assessment of the significance of the effect. 

To provide more efficient estimates as well as to control for potential confounding, I also run a set of 2SLS models. Without loss of generality, each model is presented as a system of two linear models:

\begin{equation}
    D_i = \alpha_0 + \mathbf{Z}_i' \nu + \mathbf{X}_i' \gamma + \xi_i
    \label{eq:2sls_1}
\end{equation}
\begin{equation}
    Y_i = \beta_0 + \tau \hat D_i + \mathbf{X}_i' \theta + \varepsilon_i
    \label{eq:2sls_2}
\end{equation}
where $\mathbf{Z}_i$ is a vector of instrumental variables, $\mathbf{X}_i$ is a vector of control variables and $\hat D_i$ is a prediction from \eqref{eq:2sls_1} that is a "cleaned" from endogenous variation version of observed potential outcome $D_i$. Therefore, given exogeneity of $\mathbf{Z}_i$ and $\mathbf{X}_i$, as well as $\mathbf{\nu} \neq \mathbf{0}$,  $\hat D_i$ is exogenous and $Cov(\hat D_i, \varepsilon_i) = 0$, meaning it is as good as randomly assign. Thus, $\hat \tau_{2SLS}$ from \eqref{eq:2sls_2} is consistent estimate of $\tau$.

As a first instrument for \eqref{eq:2sls_1}, I use a binary variable for same-sex: $Z^{samesex}_i = 1$ if $s_{1i} = s_{2i}$, where $s_{1i}$ and $s_{2i}$ is a sex of a first and a second child respectively. However, as was shown by \cite{angrist_orig}, same sex instrument is associated with sexes of children, which might affect economic behavior for reasons other than family size. Because of that sexes of children are included in the model as control variables:

\begin{equation}
    D_i = \alpha_0 + \nu z^{samesex}_i + \alpha_1 s_{1i} + \alpha_2 s_{2i} + \mathbf{X}_i' \gamma + \xi_i
    \label{eq:2sls_1_samesex}
\end{equation}
\begin{equation}
    Y_i = \beta_0 + \tau \hat D_i + \beta_1 s_{1i} + \beta_2 s_{2i} + \mathbf{X}_i' \theta + \varepsilon_i
    \label{eq:2sls_2_samesex}
\end{equation}
where $s_1$ and $s_2$ are binary variables for sex of the first and the second child respectively. 

The second set of instruments decompose same sex instrument into two: two boys and two girls. In that case, the first-stage model can be formulated as follows:
\begin{equation}
    D_i = \alpha_0 + \nu_1 z^{boys}_i + \nu_2 z^{girls}_i + \alpha_1 s_{2i} + \mathbf{X}_i' \gamma + \xi_i
    \label{eq:2sls_1_bothsex}
\end{equation}
the second-stage equation remains the same as in \eqref{eq:2sls_2_samesex}, but one of the children sex variables -- $s_{1i}$ and $s_{2i}$ -- should be omitted since instruments are just products of them. 

The third instrument is a multiple births occurrence, and the first-stage model is:
\begin{equation}
    D_i = \alpha_0 + \nu_1 z^{mltbrth}_i + \alpha_1 s_{1i} +  \alpha_2 s_{2i} + \mathbf{X}_i' \gamma + \xi_i
    \label{eq:2sls_1_multy}
\end{equation}

Finally, I estimate the over-identified 2SLS model, using same sex and multiple births to reach more precise estimates. However, such a strategy may lead to aggregation bias since two instrumental variables -- siblings sex composition and multiple births -- show different sides of fertility effects on economic activity. In case of multiple births, it is a "fertility shock", since instead of one child parents obtain two at once. Obviously, the effect of such an event should be stronger than the birth of a third child after some lag from the second one. 

While Hypotheses 1-2 are straightforward and can be tested by usage of different specifications of 2SLS models, Hypothesis 3 expects heterogeneity in the effect of fertility by education of mothers. The introduction of interaction terms is a common and effective strategy; however, it is not feasible with the 2SLS model described above due to the limited number of instrumental variables. Since the effect of childbearing varies across education levels (and not necessarily in a linear fashion), one should use categorical variable for education with at least several levels. In that case the number of endogenous variables in $\mathbf{D}_i$ equals the number of levels plus 1, which requires at least the same number of instruments in $\mathbf{Z}_i$ (see \eqref{eq:2sls_1}). Therefore, I employ identical models on subsamples of mothers (and fathers) with different levels of education. This approach effectively addresses the issue of an excessive number of exogenous variables while still enabling the between-groups comparisons.

To test the relevance assumption -- the significant and strong enough link between instrument(s) and endogenous variable -- I use robust F-statistic to control for possible heterogeneity applying HC1 correction for variance-covariance matrix (the same is done for the second stage variance estimates), because F-statistic under homogeneity assumption might be inflated. Additionally, for the 2SLS model with one instrument -- same sex or multiple births -- I report Anderson-Rubin 95\% CI (also known as weak-instrument interval) that is robust to possible weakness of the instrument \citep{ar_test_1}. All that concerns on the weakness tests stems from the fact that weak instruments might be more biased than usual OLS estimates if exogeneity or exclusion restriction assumptions only partially hold \citep[e.g.][]{weak_inst_1}. 

The final point to be made is the effect estimation of childbearing on employment that is a binary dependent variable. One possible strategy is to run a probit or logistic version of instrumental variable regression. However, results from linear 2SLS model and other instrumental variables estimators that explicitly assume nonlinear data-generating process do not differ significantly \citep{bin_iv}\footnote{What is more, \cite{bin_iv} demonstrates that replicating the paper by \cite{angrist_orig} -- exactly the same research design on the effect of fertility on labor market outcomes that is used in the presented paper.}. Hence, I choose to exploit linear 2SLS estimator as an optimal model with better and simpler properties. 

\subsubsection{Panel matching}

As an estimator for difference-in-difference, panel matching \citep{panelmatch} is used that combines usual DiD with matching methods for panel data. It, firstly, matched individuals by their treatment histories in a respective time $t$ (that controls for shared time characteristics and underlying life-course), ensuring treatment and control groups differ only by treatment in $t$. Consequently, matching on covariates is performed to address selection into treatment. As was previously stated, DiD strategy relaxes the strict exogeneity assumption by relying on sequential ignorability, a concept which is nearly equivalent to PTA. This can be partially tested empirically, with pre-treatment trajectories in dependent and control variables required to be similar in the treated and control groups. The existence of such a similarity implies that, in the presence of time-varying unobserved confounders, past treatment history, shared similarity in covariates' trajectories, and exact match on time will reach conditional PTA and make causal identification possible.

Thus, the primary decision for the researcher to make is how to effectively control for past to get desired parallel trends. The first issue when dealing with longitudinal data in the context of studying life-course events pertains to the operationalization of time. The conventional approach in  panel data studies involves treating the year of survey (wave) as such. Nevertheless, I consider this dimension to be irrelevant in the context of controlling for underlying life-course events between the control and treated groups. Consequently, the present study employs age as a time dimension, facilitating the comparison of treated and control women within the same age. Secondly, the number of pre-treated periods to control for is also a decision which is open to ambiguity. On the one hand, a longer pre-treatment period will facilitate the identification of the most appropriate control group. However, on the other hand, it will significantly reduce the sample size (note, the RLMS is a highly imbalanced data) and the quality of the control group, since there will be fewer individuals to choose from to recreate characteristics of the treated. In line with the previous research, the current study will examine a four-year period prior to the childbearing, so it is approximately three years before the pregnancy. I consider this length to be sufficient to verify the PTA. 

In addition, the present study employs a subsample analysis, with the sample divided into two distinct groups: firstly, women with partners, and secondly, employed women throughout the entire period under investigation. In line with Hypotheses and other expectations, the effect of childbearing on employment, hours worked per week, monthly income and job satisfaction is presented in this study. These variables are treated as both controls and dependent variables, because PTA requires the construction of groups based on past outcome histories. The description of control and dependent variables is presented above in Section \ref{sec:rlms_data}. 

To decompose the impact of childbearing across parities, the following comparisons are employed: between women who have never given birth and those who have given birth to their first child; between women with one child and women who have given birth to their second child; and finally, between women with two children and women who have given birth to their third child. Thus, only the group of mothers with one child is compared to childless women, while other groups with two or three children are compared to mothers of lower parity.

The final technical points should be made. Firstly, I match upon covariate histories prior to birth and pregnancy to assure that controls are not subject to posttreatment bias. Secondly, to calculate variance of dynamic ATT, block bootstrap is used and the percentile CI is shown, which is robust to possible asymmetry of estimates distribution (though they are more conservative). Finally, to balance treated and control groups upon covariates, the Covariate Balancing Propensity Score (CBPS) is used \citep[see][]{cbps}.

\subsection{Data \& Variables}\label{sec:data}

\subsubsection{Cross-sectional data}

As a main source of data for IV analysis I use 10\% random samples from Russian Censuses of 2002 and 2010 \citep{census2002, census2010} that is similar to strategy by \cite{angrist_orig}. Since census is an example of cross-sectional data, I do not study possible temporal variation of the parenthood penalty explicitly, but do independent analysis on each sample to get more robust results. Unfortunately, in censuses there is only a modest set of variables that might reveal the parenthood penalty in Russia. Hence, as main dependent variables I consider employment and the presence of a second job (the last variable is available only for 2010 census)\footnote{Precisely, "employment" variable is a response to the question asking whether the respondent had a paid job in the time of census. Concurrently, "second job" is a response to the question asking whether the respondent had a second paid job in the time of census} since there are no indicators for income, wage and working hours.

For the IV analysis, the specific sub-sample of mothers from censuses is used. The samples are limited exclusively to mothers with a minimum of two children, as the instruments employed – namely, siblings, sex composition and multiple birth – are only able to demonstrate the effect of childbearing through the comparison of mothers with a two or more children. Secondly, the anonymity of census data ensures that it is not possible to identify separated children. Consequently, the sample is limited to mothers aged 18-55 whose eldest child is under the age of 18, what ensures that probability of moving to another household is small. To estimate the effect of childbearing on husbands, only one additional restriction to the previous sub-samples is needed -- by marital status. In other words, the husbands' analysis exploits sample of married mothers to whom husbands characteristics were matched, whereby it restricts husbands' sample only to cohabit couples. Concurrently, there is no restriction on formal marriage and couples can be unregistered.

As control variables from the censuses following personal characteristics are collected: age, type of place of living (rural/urban), region of living and education. To the model specifications in \eqref{eq:2sls_1_samesex}-\eqref{eq:2sls_1_multy} as controls (matrix $\mathbf{X}_i'$) the following variables are included: mother's age (categorical variable with 5-year age groups where "<25" and "45+" are lowest and highest categories respectively), children's age (treated as continuous variable), marital status (binary variable, identifying whether there is a spouse) and rural dummy as well as region FE. Education is excluded from this list since it might be endogenous: level of education is affected by fertility. However, further I present results on sub-samples by mother's education to find potential heterogeneity in the parenthood penalty. 

The main limitation with census' data is its quality. In academic community Russian censuses are considered at best as moderately acceptable \citep{census_problems1, census_problems3}\footnote{According to the results of a survey conducted by Russian Public Opinion Research Center \citep{census_problems2}, in 2010 census 11\% of respondents were not enumerated in the census, while 22\% were enumerated by relatives. In turn, in 2002 census 5\% and 19\% were not enumerated at all or were enumerated by relatives respectively.}. Furthermore, the issue of underestimation of the population is especially pronounced for children under age 5, who partially form the sample of mothers in the presented study. Thus, the results might be not representative; and, secondly, biased due to "death souls" -- fictitious individuals who emerged to fulfill census enumerators' quotas, -- undercounting of individuals in specific age groups, and double counting, when one individual might be enumerated twice (though the latter problem is primarily common for internal migrants). The first limitation is not severe since the sample is large enough to consider it as an interesting population of russians who appears in 2002 or 2010 censuses. The second limitation, in turn, might lead to selection bias because the sample may have excluded a significant proportion of mothers with the youngest children. 

\begin{figure}[h] 
    \begin{minipage}{1\textwidth}
        \centering
        \includegraphics[width=0.8\textwidth]{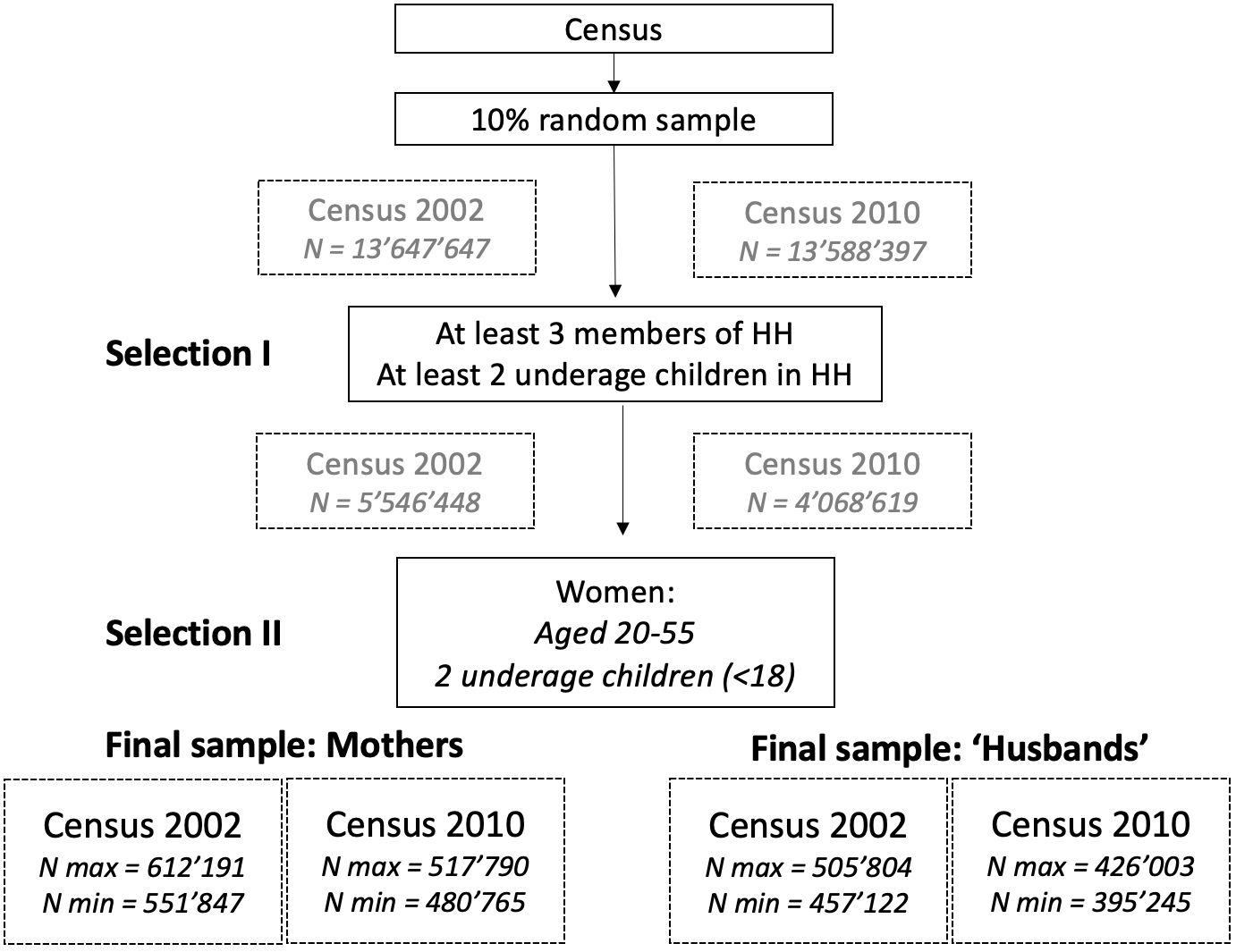}
        \caption{Description of the construction of final datasets from the original censuses of 2002 and 2010.}
        \label{fig:desc_census_selection}
    \end{minipage}\hfill
    \small{\textit{Note: N max indicates the maximum number of observations without taking into account missing values in the control variables, while N min indicates the number of observations when removing missing values for all variables used.}}
\end{figure}

In summary, the final census samples comprise mothers aged between 20 and 55 with a minimum of two children, with the eldest child being under the age of 18. The descriptive statistics for the dependent, independent and control variables by instruments' levels are presented in the Appendix \ref{app:census}. The final samples are comprised of 553,394 and 482,260 mothers for the 2002 and 2010 censuses respectively, and 458,359 and 396,460 fathers for the same years respectively\footnote{These numbers correspond to samples after listwise deletion. I assume that missings are distributed at random. Nevertheless, IV approach works well in case of violation of that assumption, and the estimates still have desired causal properties.}. The selection procedure is summarized in Figure \ref{fig:desc_census_selection}. 

\subsubsection{Longitudinal data}\label{sec:rlms_data}

As a source for longitudinal data I use Russian Longitudinal Monitoring Survey (RLMS, \cite{rlms}), where households are monitored across annual waves. However, this longitudinal dimension is not representative because of randomization design of the survey that assure only cross-sectional representation of russian population\footnote{See \cite{rlms_desc} for a brief description of RLMS data.}. Nevertheless, the number of monitored households is large enough and is rising from wave to wave to consider this data source as being of interest to the study. In contrast to censuses described above, in RLMS there are a plethora of variables that are considered as dependent ones in the research. Namely, employment, second job, average wage, income, working hours per weak, subjective health and life-satisfaction. 

The set of variables used for matching are: year of birth to control for potential heterogeneity between women of the same age, but from different cohorts; binary variable for graduation; employment\footnote{The survey question (j1) is: "Your main occupation at present. Are you currently employed, are you on paid or unpaid leave, or are you unemployed?". I operationalize employment (=1) as the answer "You're working right now", while considering other options as "0".}; 
number of worked hours per week (unemployed women get 0)\footnote{The survey question (j6.2) is: "How many hours on average does your typical work week last?". There are some individuals who reported to work 168 hours per week (meaning 24 hours per day). I consider such answers as mistakes and set the maximum number of hours worked to 112 (16 hours per day, including Saturday and Sunday).}; 
income per month (in 2022 rubles, logarithmic)\footnote{The survey question (j60) is: "How much money have you personally received in the last 30 days, counting everything: salary, pensions, bonuses, profits, allowances, financial aid, casual earnings, and other cash receipts?". The income rather than wage is used to account for the government's financial contribution. Furthermore, RLMS provides solely the respondents' answers to financial questions; consequently, the data reflects current prices. To facilitate the comparison, I convert income to constant 2022 rubles using Rosstat data on the annual consumer prices inflation \citep{infl}.}; 
subjective life satisfaction (categorical variable from 1 to 5 that is treated in the analysis as continuous variable)\footnote{The survey question (j65) is: "How satisfied are you with your life in general at present?", where 1 is fully satisfied and 5 is fully dissatisfied};
binary variable for partnership (both registered and not) as well dummy variable for urban placement. 

In the case of the sample comprised of women with partners, the following characteristics of their partners were added as control variables: graduation, employment status, the number of hours worked per week, and monthly income. In the case of the sample of employed women, job satisfaction (categorical variable from 1 to 5 that is treated in the analysis as continuous variable)\footnote{The survey question (j1.1.1) is: "Is the state the owner or co-owner of your enterprise, organization?"} 
and the dummy for public sector employment\footnote{The survey question (j23) is: "Please tell me, how satisfied or dissatisfied are you with your job in general?", where 1 is fully satisfied and 5 is fully dissatisfied} 
are added as additional control variables. 

In contrast to censuses, RLMS data is collected by professional enumerators that decreases significantly bias due to "death souls" as well as drops potential measurement error due to mistakes in coding. Descriptive statistic for dependent and some other variables are presented in Appendix \ref{app:rlms}.

\section{Results}\label{sec:results}

This section is devoted to the study's findings. Firstly, in Section \ref{sec:results_census} the primary results from the 2002 and 2010 censuses and instrumental variables usage are examined. In the following Section \ref{sec:results_rlms}, further results derived from the RLMS data with panel matching are presented. Later results are employed as an auxiliary resource and validation to the results derived from the census data.

\subsection{Results with census data}\label{sec:results_census}

In Table \ref{tab:wald_est}, the Wald estimates are presented. The first three columns report the instrument name, share of women with 3 children (treated group) and share of women with the value of instrument equal to "1". Subsequent three columns are devoted for calculation of compliers populations. It is seen that in case of sibling's instruments, the share of compliers is relatively small: only around 3\% of individuals with the same sex of children opted for a subsequent child. Interestingly, but these shares do not differ 2002 and 2010 censuses, suggesting the independence of instruments from temporal effects. In contrast, in case of multiple births the share of compliers is approaching 100 percent, though the share of women who experienced multiple births is small -- just 1 percent for both 2002 and 2010 data. The Wald estimates for the 2002 data are controversial and insignificant, suggesting the absence of childbearing effect on mother's employment. In contrast, the Wald estimates for the 2010 data depicts strong and negative effect of childbearing on labor market outcome. Same sex instrument suggests that permanent effect of childbearing on employment is about 10\%, while multiple birth shows 6\% decrease. As was outlined in Section \ref{sec:2sls}, Wald estimate, firstly, fail to adjust for individual's characteristics, and secondly, are usually uncertain that is, indeed, the case in Table \ref{tab:wald_est}.

\begin{ThreePartTable}
\begin{TableNotes}[para]
\item \textit{\scriptsize Note:} 
\item \begingroup \scriptsize $\hat \tau_{Wald}$ shows Wald estimates calculated as in \eqref{eq:ivwald_theory}; $SE(\hat \tau_{Wald})$ is standard error calculated by nonparametric bootstrap with 500 samples. The first stage is the denominator of \eqref{eq:ivwald_theory} which shows the total compliers' share in the population; the share of compliers in the treated population is $\mathbb{P}(D^{Z=1} > D^{Z=0}|D=1)$, while the share of compliers in the untreated population is $\mathbb{P}(D^{Z=1} > D^{Z=0}|D=0)$. \endgroup
\end{TableNotes}
\begin{longtable}[t]{>{\raggedright\arraybackslash}p{2cm}>{\centering\arraybackslash}p{1.1cm}>{\centering\arraybackslash}p{1.1cm}>{\centering\arraybackslash}p{2.2cm}>{\centering\arraybackslash}p{2.2cm}>{\centering\arraybackslash}p{2.2cm}>{\centering\arraybackslash}p{1.5cm}>{\centering\arraybackslash}p{1.5cm}}
\caption{\label{tab:wald_est}Wald estimates of the effect of chilbearing on mother's employment and complience probabilities, census data 2002 and 2010.}\\
\toprule
IV & $\mathbb{P}(D)$ & $\mathbb{P}(Z)$ & \shortstack[l]{1st stage\\ (compliers)} & \shortstack[l]{Compliers\\ $D=1$} & \shortstack[l]{Compliers\\ $D=0$} & $\hat \tau_{Wald}$ & $SE(\hat \tau_{Wald})$\\
\midrule[1pt] \multicolumn{8}{c}{\textbf{Census 2002}} \\ \midrule[1pt]
\midrule
Same sex & 0.16 & 0.51 & 0.03 & 0.08 & 0.02 & 0.01 & 0.11\\
\addlinespace[0.5em]
Both boys & 0.16 & 0.26 & 0.01 & 0.02 & 0.01 & -0.08 & 0.41\\
\addlinespace[0.5em]
Both girls & 0.16 & 0.24 & 0.02 & 0.04 & 0.02 & 0.05 & 0.14\\
\addlinespace[0.5em]
Multiple births & 0.16 & 0.01 & 0.85 & 0.05 & 1.00 & 0.00 & 0.02\\
\midrule[1pt] \multicolumn{8}{c}{\textbf{Census 2010}} \\ \midrule[1pt]\\
\addlinespace[0.5em]
Same sex & 0.17 & 0.50 & 0.03 & 0.08 & 0.02 & -0.10 & 0.05\\
\addlinespace[0.5em]
Both boys & 0.17 & 0.27 & 0.01 & 0.01 & 0.01 & -0.46 & 0.19\\
\addlinespace[0.5em]
Both girls & 0.17 & 0.24 & 0.03 & 0.04 & 0.03 & 0.02 & 0.06\\
\addlinespace[0.5em]
Multiple births & 0.16 & 0.01 & 0.84 & 0.04 & 1.00 & -0.06 & 0.01\\
\bottomrule
\insertTableNotes
\end{longtable}
\end{ThreePartTable}

In turn, in Table \ref{tab:census2002_mothers_nocontrols} and Table \ref{tab:census2010_mothers_nocontrols} the analysis of the effect of childbearing on mother's employment is presented with the usage of 2002 or 2010 censuses respectively. In each Table the first model corresponds to OLS estimation that shows highly negative and significant effect of fertility on labor supply: the presence of the third child, ceteris paribus, reduces employment by about 20 percentage points (0.17 and 0.20 for 2002 and 2010 census data respectively). In contrast, 2SLS estimates of the effect differ dramatically from OLS results, with only the model with multiple births displaying a negative effect of approximately 6 percentage points. Conversely, models employing same-sex instruments fail to identify any significant effects associated with the third child. Given the sufficient strength of the instruments (all tests on weakness are rejected at any reasonable significance level), the analysis suggests that OLS estimates are considerably inflated, while the causal effect is modest. However, as was previously stated, the multiple births instrument demonstrates the impact of fertility shock. Conversely, the same-sex instruments identify the general effect of fertility, which, as proposed by the analysis, is negligible.

\begin{table}[H]
\caption{OLS and 2SLS estimates of the effect of childbearing on mother's employment, 2002 census.}
\begin{center}
\begin{threeparttable}
\begin{tabular}{l D{.}{.}{6.5} D{.}{.}{6.9} D{.}{.}{6.6} D{.}{.}{6.12}}
\toprule
 & \multicolumn{1}{c}{OLS} & \multicolumn{1}{c}{2SLS} & \multicolumn{1}{c}{2SLS} & \multicolumn{1}{c}{2SLS} \\
\midrule
Intercept            & 0.26^{***}                     & 0.27^{***}                           & 0.26^{***}                                  & 0.27^{***}                                  \\
                     & (0.01)                         & (0.01)                               & (0.01)                                      & (0.01)                                      \\
More than 2 children & -0.17^{***}                    & -0.04                                & -0.02                                       & -0.03^{***}                                 \\
                     & (0.00)                         & (0.05)                               & (0.04)                                      & (0.01)                                      \\
1st child sex        & -0.00^{***}                    & -0.00^{***}                          &                                             & -0.00^{***}                                 \\
                     & (0.00)                         & (0.00)                               &                                             & (0.00)                                      \\
2nd child sex        & -0.01^{***}                    & -0.01^{***}                          & -0.01^{***}                                 & -0.01^{***}                                 \\
                     & (0.00)                         & (0.00)                               & (0.00)                                      & (0.00)                                      \\
\midrule
IV                   & \multicolumn{1}{c}{\small{--}} & \multicolumn{1}{c}{\small{Same sex}} & \multicolumn{1}{c}{\small{Both boys/girls}} & \multicolumn{1}{c}{\small{Multiple births}} \\
Weak Instr.          & \multicolumn{1}{c}{--}         & \multicolumn{1}{c}{939.34 ***}       & \multicolumn{1}{c}{479.79 ***}              & \multicolumn{1}{c}{33059.26 ***}            \\
AR                   & \multicolumn{1}{c}{--}         & \multicolumn{1}{c}{(-0.13) - 0.05}   & \multicolumn{1}{c}{--}                      & \multicolumn{1}{c}{(-0.05) - (-0.02)}       \\
Wu-Hausman           & \multicolumn{1}{c}{--}         & \multicolumn{1}{c}{8.04 **}          & \multicolumn{1}{c}{11.31 ***}               & \multicolumn{1}{c}{325.44 ***}              \\
Sargan               & \multicolumn{1}{c}{--}         & \multicolumn{1}{c}{--}               & \multicolumn{1}{c}{11.4 ***}                & \multicolumn{1}{c}{--}                      \\
Num. obs.            & 553394                         & 553394                               & 553394                                      & 551847                                      \\
\bottomrule
\end{tabular}
\begin{tablenotes}[flushleft]
\scriptsize{\item Note: $^{***}p<0.001$; $^{**}p<0.01$; $^{*}p<0.05$. As control variables in all models are included mother's age (categorical variable with 5-year age groups), children's age, marital status and rural dummy as well as region FE; heteroskedasticity consistent standard errors (HC1) are in parentheses; test on weak instruments shows robust F-statistic from the first-stage; AR - Anderson-Rubin 95-CI.}
\end{tablenotes}
\end{threeparttable}
\label{tab:census2002_mothers_nocontrols}
\end{center}
\end{table}

\begin{table}[H]
\caption{OLS and 2SLS estimates of the effect of childbearing on mother's employment, 2010 census.}
\begin{center}
\begin{threeparttable}
\begin{tabular}{l D{.}{.}{6.5} D{.}{.}{6.9} D{.}{.}{6.6} D{.}{.}{6.12}}
\toprule
 & \multicolumn{1}{c}{OLS} & \multicolumn{1}{c}{2SLS} & \multicolumn{1}{c}{2SLS} & \multicolumn{1}{c}{2SLS} \\
\midrule
Intercept            & 0.34^{***}                     & 0.35^{***}                           & 0.35^{***}                                  & 0.36^{***}                                  \\
                     & (0.02)                         & (0.02)                               & (0.02)                                      & (0.02)                                      \\
More than 2 children & -0.20^{***}                    & -0.08                                & -0.05                                       & -0.03^{***}                                 \\
                     & (0.00)                         & (0.05)                               & (0.05)                                      & (0.01)                                      \\
1st child sex        & -0.00^{**}                     & -0.00^{*}                            &                                             & -0.00^{*}                                   \\
                     & (0.00)                         & (0.00)                               &                                             & (0.00)                                      \\
2nd child sex        & -0.01^{***}                    & -0.01^{***}                          & -0.01^{***}                                 & -0.01^{***}                                 \\
                     & (0.00)                         & (0.00)                               & (0.00)                                      & (0.00)                                      \\
\midrule
IV                   & \multicolumn{1}{c}{\small{--}} & \multicolumn{1}{c}{\small{Same sex}} & \multicolumn{1}{c}{\small{Both boys/girls}} & \multicolumn{1}{c}{\small{Multiple births}} \\
Weak Instr.          & \multicolumn{1}{c}{--}         & \multicolumn{1}{c}{655.97 ***}       & \multicolumn{1}{c}{343.46 ***}              & \multicolumn{1}{c}{29707.43 ***}            \\
AR                   & \multicolumn{1}{c}{--}         & \multicolumn{1}{c}{(-0.18) - 0.02}   & \multicolumn{1}{c}{--}                      & \multicolumn{1}{c}{(-0.05) - (-0.02)}       \\
Wu-Hausman           & \multicolumn{1}{c}{--}         & \multicolumn{1}{c}{5.16 *}           & \multicolumn{1}{c}{8.33 **}                 & \multicolumn{1}{c}{448.37 ***}              \\
Sargan               & \multicolumn{1}{c}{--}         & \multicolumn{1}{c}{--}               & \multicolumn{1}{c}{6.6 *}                   & \multicolumn{1}{c}{--}                      \\
Num. obs.            & 482260                         & 482260                               & 482260                                      & 480765                                      \\
\bottomrule
\end{tabular}
\begin{tablenotes}[flushleft]
\scriptsize{\item Note: $^{***}p<0.001$; $^{**}p<0.01$; $^{*}p<0.05$. As control variables in all models are included mother's age (categorical variable with 5-year age groups), children's age, marital status and rural dummy as well as region FE; heteroskedasticity consistent standard errors (HC1) are in parentheses; test on weak instruments shows robust F-statistic from the first-stage; AR - Anderson-Rubin 95-CI.}
\end{tablenotes}
\end{threeparttable}
\label{tab:census2010_mothers_nocontrols}
\end{center}
\end{table}

In Table \ref{tab:census2002_fathers_nocontrols} and Table \ref{tab:census2010_fathers_nocontrols} the analysis of the effect of childbearing on husband's employment is presented with the usage of 2002 or 2010 censuses respectively. All models are analogous to aforementioned in Table \ref{tab:census2002_mothers_nocontrols} and Table \ref{tab:census2010_mothers_nocontrols} for mother's employment. Similar to motherhood analysis, OLS estimates report significantly negative results: the presence of the third child reduces employment by around 3 percentage points. Meanwhile, all 2SLS models identify no significant negative effect of parenthood as well as any "premia" that is in line with Hypothesis 2.

\begin{table}[H]
\caption{OLS and 2SLS estimates of the effect of childbearing on husband's employment, 2002 census.}
\begin{center}
\begin{threeparttable}
\begin{tabular}{l D{.}{.}{6.5} D{.}{.}{6.9} D{.}{.}{6.6} D{.}{.}{6.9}}
\toprule
 & \multicolumn{1}{c}{OLS} & \multicolumn{1}{c}{2SLS} & \multicolumn{1}{c}{2SLS} & \multicolumn{1}{c}{2SLS} \\
\midrule
Intercept            & 0.76^{***}                     & 0.76^{***}                           & 0.76^{***}                                  & 0.76^{***}                                  \\
                     & (0.04)                         & (0.04)                               & (0.04)                                      & (0.04)                                      \\
More than 2 children & -0.03^{***}                    & 0.02                                 & 0.02                                        & 0.00                                        \\
                     & (0.00)                         & (0.04)                               & (0.04)                                      & (0.01)                                      \\
1st child sex        & 0.00                           & 0.00                                 &                                             & 0.00                                        \\
                     & (0.00)                         & (0.00)                               &                                             & (0.00)                                      \\
2nd child sex        & -0.00                          & -0.00                                & -0.00                                       & -0.00                                       \\
                     & (0.00)                         & (0.00)                               & (0.00)                                      & (0.00)                                      \\
\midrule
IV                   & \multicolumn{1}{c}{\small{--}} & \multicolumn{1}{c}{\small{Same sex}} & \multicolumn{1}{c}{\small{Both boys/girls}} & \multicolumn{1}{c}{\small{Multiple births}} \\
Weak Instr.          & \multicolumn{1}{c}{--}         & \multicolumn{1}{c}{959.78 ***}       & \multicolumn{1}{c}{489.14 ***}              & \multicolumn{1}{c}{26866.29 ***}            \\
AR                   & \multicolumn{1}{c}{--}         & \multicolumn{1}{c}{(-0.05) - 0.09}   & \multicolumn{1}{c}{--}                      & \multicolumn{1}{c}{(-0.01) - 0.02}          \\
Wu-Hausman           & \multicolumn{1}{c}{--}         & \multicolumn{1}{c}{2.36  }           & \multicolumn{1}{c}{2.26  }                  & \multicolumn{1}{c}{26.66 ***}               \\
Sargan               & \multicolumn{1}{c}{--}         & \multicolumn{1}{c}{--}               & \multicolumn{1}{c}{0.11  }                  & \multicolumn{1}{c}{--}                      \\
Num. obs.            & 458359                         & 458359                               & 458359                                      & 457122                                      \\
\bottomrule
\end{tabular}
\begin{tablenotes}[flushleft]
\scriptsize{\item Note: $^{***}p<0.001$; $^{**}p<0.01$; $^{*}p<0.05$. As control variables in all models are included mother's and husband's ages (categorical variables with 5-year age groups), children's age and rural dummy as well as region FE; heteroskedasticity consistent standard errors (HC1) are in parentheses; test on weak instruments shows robust F-statistic from the first-stage; AR - Anderson-Rubin 95-CI.}
\end{tablenotes}
\end{threeparttable}
\label{tab:census2002_fathers_nocontrols}
\end{center}
\end{table}

\begin{table}[H]
\caption{OLS and 2SLS estimates of the effect of childbearing on husband's employment, 2010 census.}
\begin{center}
\begin{threeparttable}
\begin{tabular}{l D{.}{.}{6.5} D{.}{.}{6.9} D{.}{.}{6.6} D{.}{.}{6.9}}
\toprule
 & \multicolumn{1}{c}{OLS} & \multicolumn{1}{c}{2SLS} & \multicolumn{1}{c}{2SLS} & \multicolumn{1}{c}{2SLS} \\
\midrule
Intercept            & 0.74^{***}                     & 0.74^{***}                           & 0.74^{***}                                  & 0.74^{***}                                  \\
                     & (0.04)                         & (0.04)                               & (0.04)                                      & (0.04)                                      \\
More than 2 children & -0.02^{***}                    & -0.01                                & -0.01                                       & -0.00                                       \\
                     & (0.00)                         & (0.03)                               & (0.03)                                      & (0.01)                                      \\
1st child sex        & -0.00                          & -0.00                                &                                             & -0.00                                       \\
                     & (0.00)                         & (0.00)                               &                                             & (0.00)                                      \\
2nd child sex        & -0.00                          & -0.00                                & -0.00                                       & -0.00                                       \\
                     & (0.00)                         & (0.00)                               & (0.00)                                      & (0.00)                                      \\
\midrule
IV                   & \multicolumn{1}{c}{\small{--}} & \multicolumn{1}{c}{\small{Same sex}} & \multicolumn{1}{c}{\small{Both boys/girls}} & \multicolumn{1}{c}{\small{Multiple births}} \\
Weak Instr.          & \multicolumn{1}{c}{--}         & \multicolumn{1}{c}{658.61 ***}       & \multicolumn{1}{c}{346.29 ***}              & \multicolumn{1}{c}{23965.06 ***}            \\
AR                   & \multicolumn{1}{c}{--}         & \multicolumn{1}{c}{(-0.08) - 0.05}   & \multicolumn{1}{c}{--}                      & \multicolumn{1}{c}{(-0.01) - 0.01}          \\
Wu-Hausman           & \multicolumn{1}{c}{--}         & \multicolumn{1}{c}{0.07  }           & \multicolumn{1}{c}{0.23  }                  & \multicolumn{1}{c}{14.54 ***}               \\
Sargan               & \multicolumn{1}{c}{--}         & \multicolumn{1}{c}{--}               & \multicolumn{1}{c}{0.83  }                  & \multicolumn{1}{c}{--}                      \\
Num. obs.            & 396460                         & 396460                               & 396460                                      & 395245                                      \\
\bottomrule
\end{tabular}
\begin{tablenotes}[flushleft]
\scriptsize{\item Note: $^{***}p<0.001$; $^{**}p<0.01$; $^{*}p<0.05$. As control variables in all models are included mother's and husband's ages (categorical variables with 5-year age groups), children's age and rural dummy as well as region FE; heteroskedasticity consistent standard errors (HC1) are in parentheses; test on weak instruments shows robust F-statistic from the first-stage; AR - Anderson-Rubin 95-CI.}
\end{tablenotes}
\end{threeparttable}
\label{tab:census2010_fathers_nocontrols}
\end{center}
\end{table}

\begin{figure}[H] 
    \begin{minipage}{1\textwidth}
        \centering
        \includegraphics[width=0.8\textwidth]{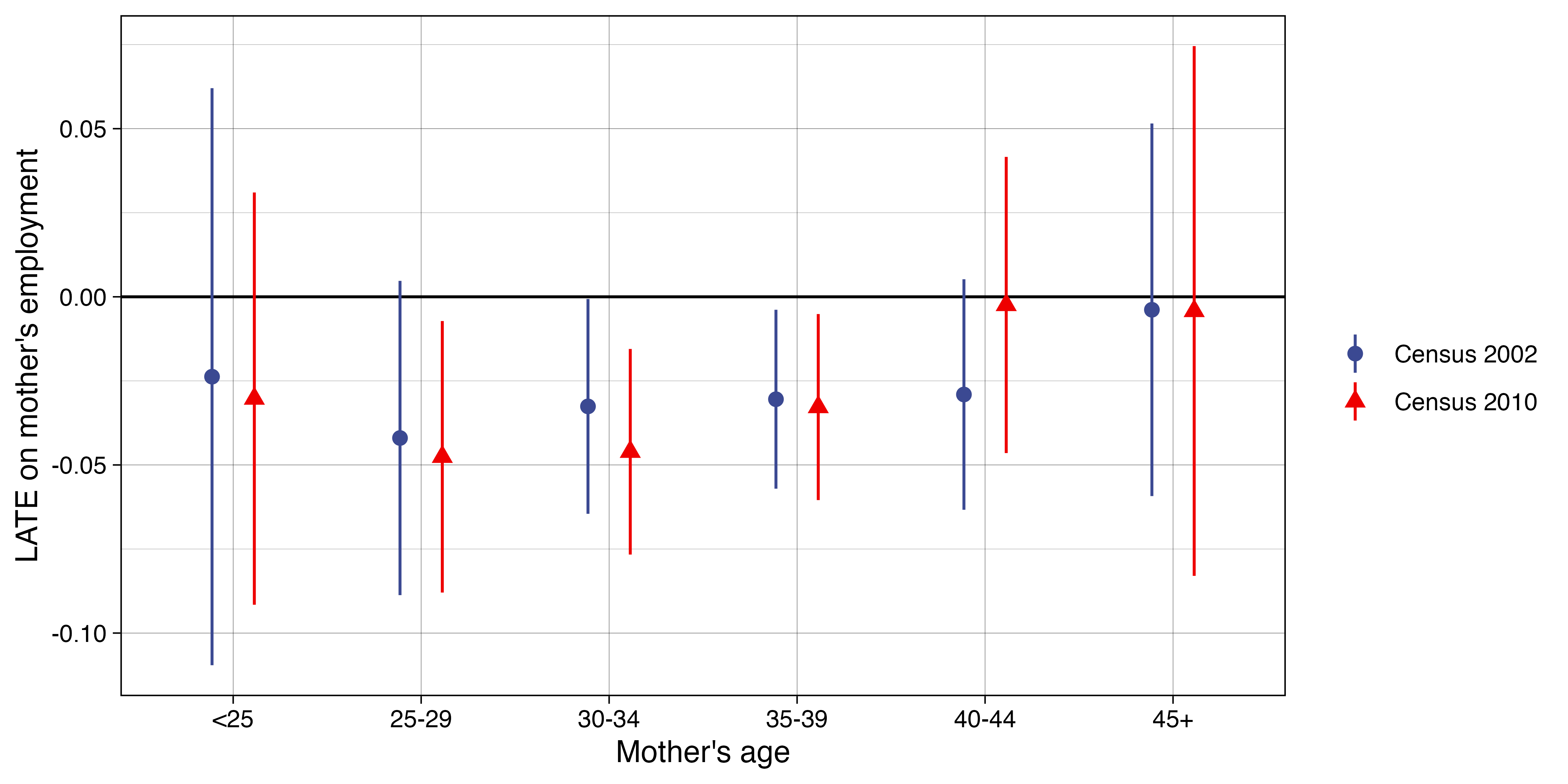}
        \caption{LATE of childbearing on mother's and husband's employment by level of education, census data.}
        \label{fig:late_on_empl_by_age_census}
    \end{minipage}\hfill
    \small{\textit{Note: Estimates are based on 2SLS models by sub-samples of the level of mother's (on the right) or husband's (on the left) education  with same sex and multiple births as instruments; covariates are mother's and father's ages (categorical variable with 5-year age groups), children's age and children's sex; lines depict 95\% confidence interval constructed using heteroskedasticity consistent standard errors (HC1).}}
\end{figure}

Next, in Figure \ref{fig:late_on_empl_by_age_census} the LATE estimates with usage of same sex and multiple birth instruments, that are employed simultaneously for distinct subsamples of mothers from various age groups. An analysis reveals significant heterogeneity by age. The most substantial penalty is observed among the 25-39 age group of mothers, while the impact on the 40+ age group remains negligible and less pronounced. Conversely, the graph demonstrates an absence of effect on the youngest group of mothers, defined as those under the age of 25. It can be explained by the fact that a proportion of women who already have at least two children at such a young age is small that do not allow for the identification of the effect. Meanwhile, the point estimates remain lower than for other age groups, with the exception of those aged 45 and over. This could be a consequence of the fact that such women are strongly predisposed to prioritise family over career, thus resulting in a less pronounced distinction between them.

\begin{figure}[H] 
    \begin{minipage}{1\textwidth}
        \centering
        \includegraphics[width=0.8\textwidth]{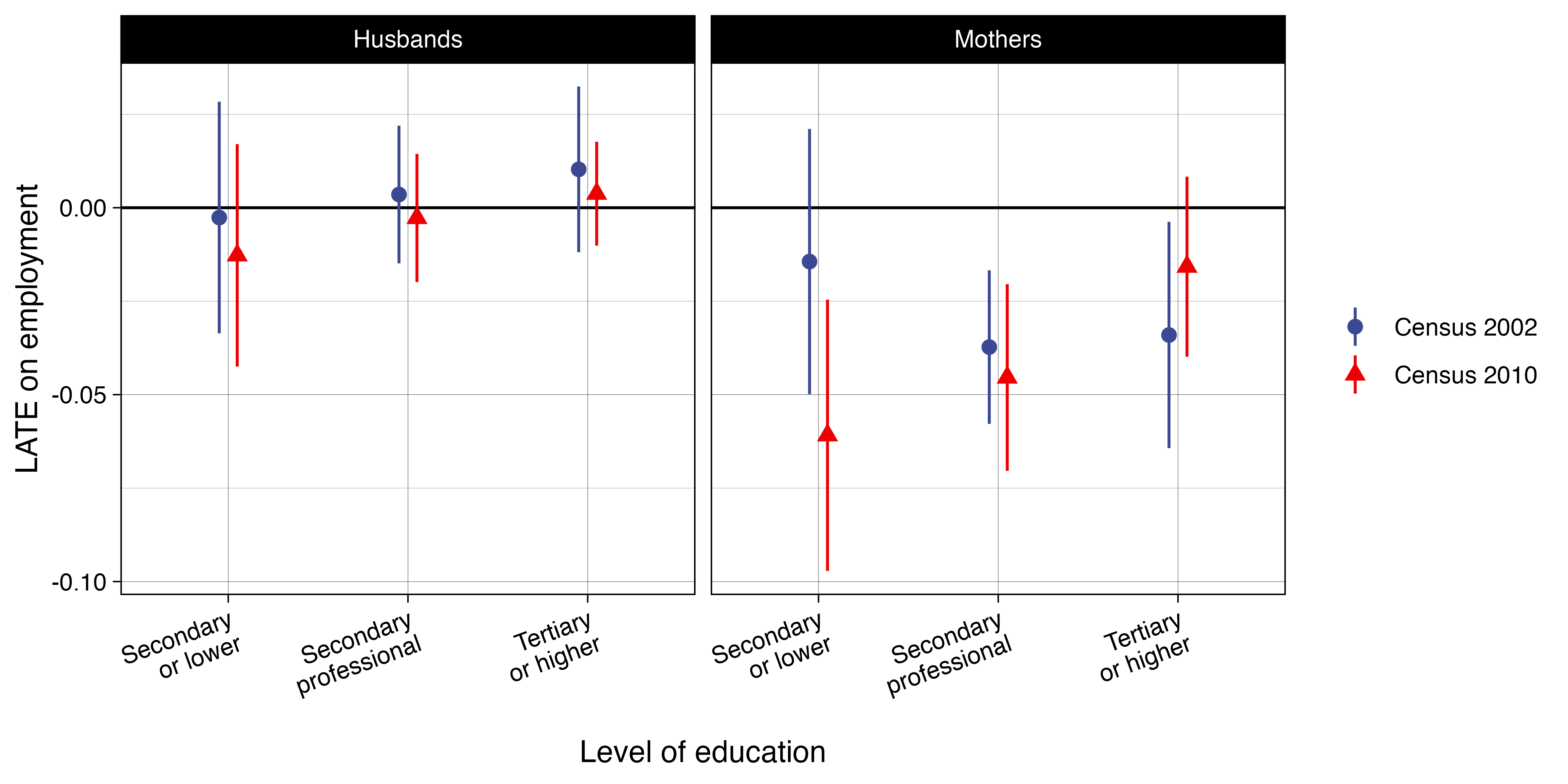}
        \caption{LATE of childbearing on mother's and husband's employment by level of education, census data.}
        \label{fig:late_on_empl_by_educ_census}
    \end{minipage}\hfill
    \small{\textit{Note: Estimates are based on 2SLS models by sub-samples of the level of mother's (on the right) or husband's (on the left) education  with same sex and multiple births as instruments; covariates are mother's and father's ages (categorical variable with 5-year age groups), children's age and children's sex; lines depict 95\% confidence interval constructed using heteroskedasticity consistent standard errors (HC1).}}
\end{figure}

In Figure \ref{fig:late_on_empl_by_educ_census} the same analysis as in Figure \ref{fig:late_on_empl_by_age_census} is presented. In this instance, however, the data is divided into subsamples according to educational attainment, categorised as secondary or lower, secondary professional, and tertiary or higher. Furthermore, the left and the right panel of the graph illustrate the impact of parenthood on father and husbands respectively. In accordance with Hypothesis 3, the motherhood penalty is most pronounced among women with low levels of education, while in the group of graduated women, the effect is indistinct from zero. Intriguingly, the analysis of the 2002 censuses also suggests the absence of the effect in the group with the lowest level of education, a phenomenon that the present study fails to explain, since it is unlikely to be a product of lack of statistical power. Nevertheless, Hypothesis 3 is supported as it is evident that there is a less pronounced penalty imposed on graduate mothers in comparison to other groups. Despite the absence of rationale to anticipate variations in the impact of fatherhood by educational levels, it remains a substantive robustness check to demonstrate that the zero effect is not attributable to the presence of heterogeneity. Indeed, the results indicate an absence of an effect of fatherhood on labor supply by education levels.

\begin{figure*}[htb]
    \centering
    \begin{subfigure}[t]{0.6\textwidth}  
        \centering
        \includegraphics[width=1\textwidth]{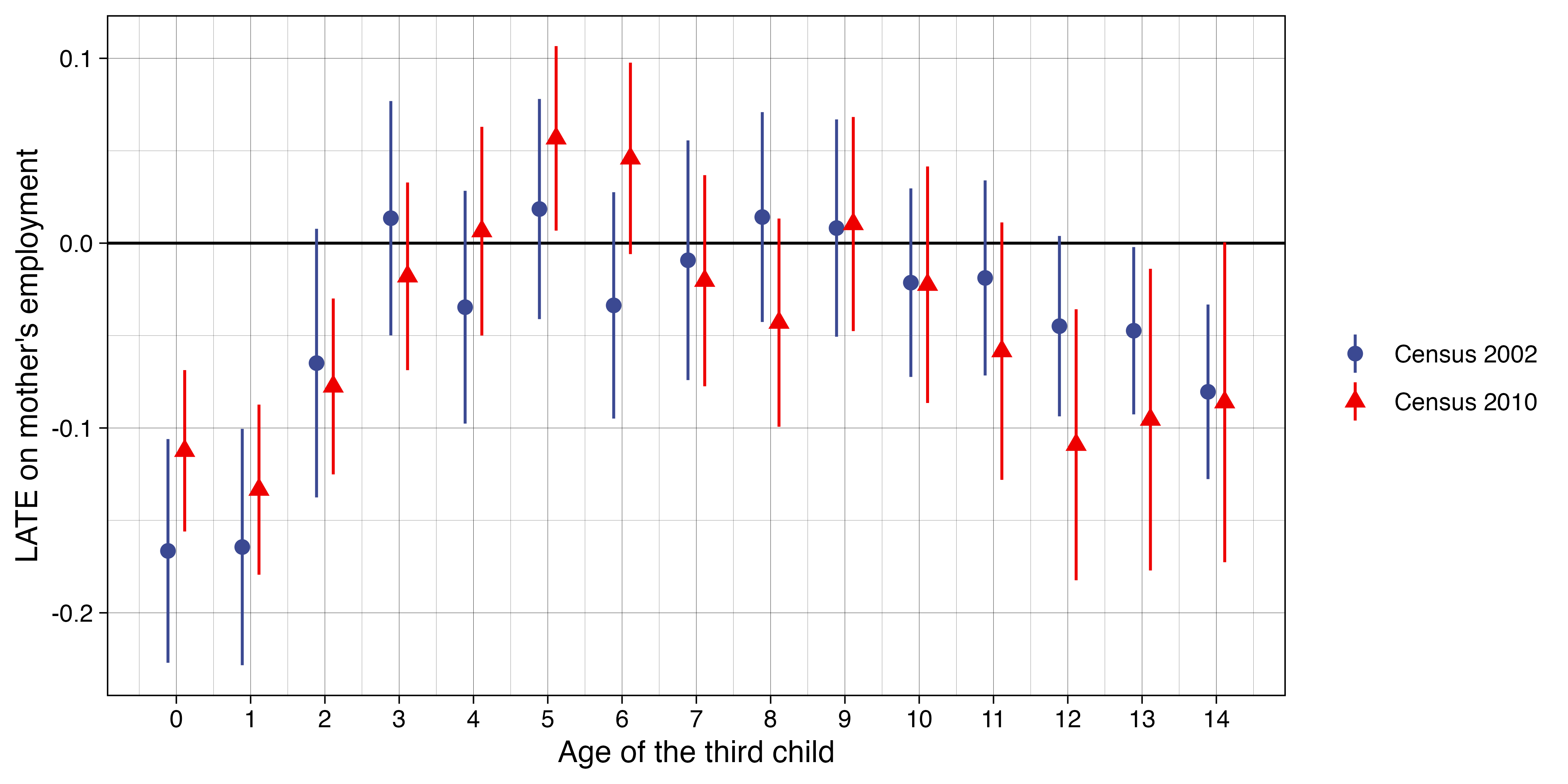}
        \caption{LATE of childbearing on mother's employment by age of the third child, census data.}
        \label{fig:late_on_empl_by_3age_census_mothers}
    \end{subfigure}
    ~
    \begin{subfigure}[t]{0.6\textwidth}
        \centering
        \includegraphics[width=1\textwidth]{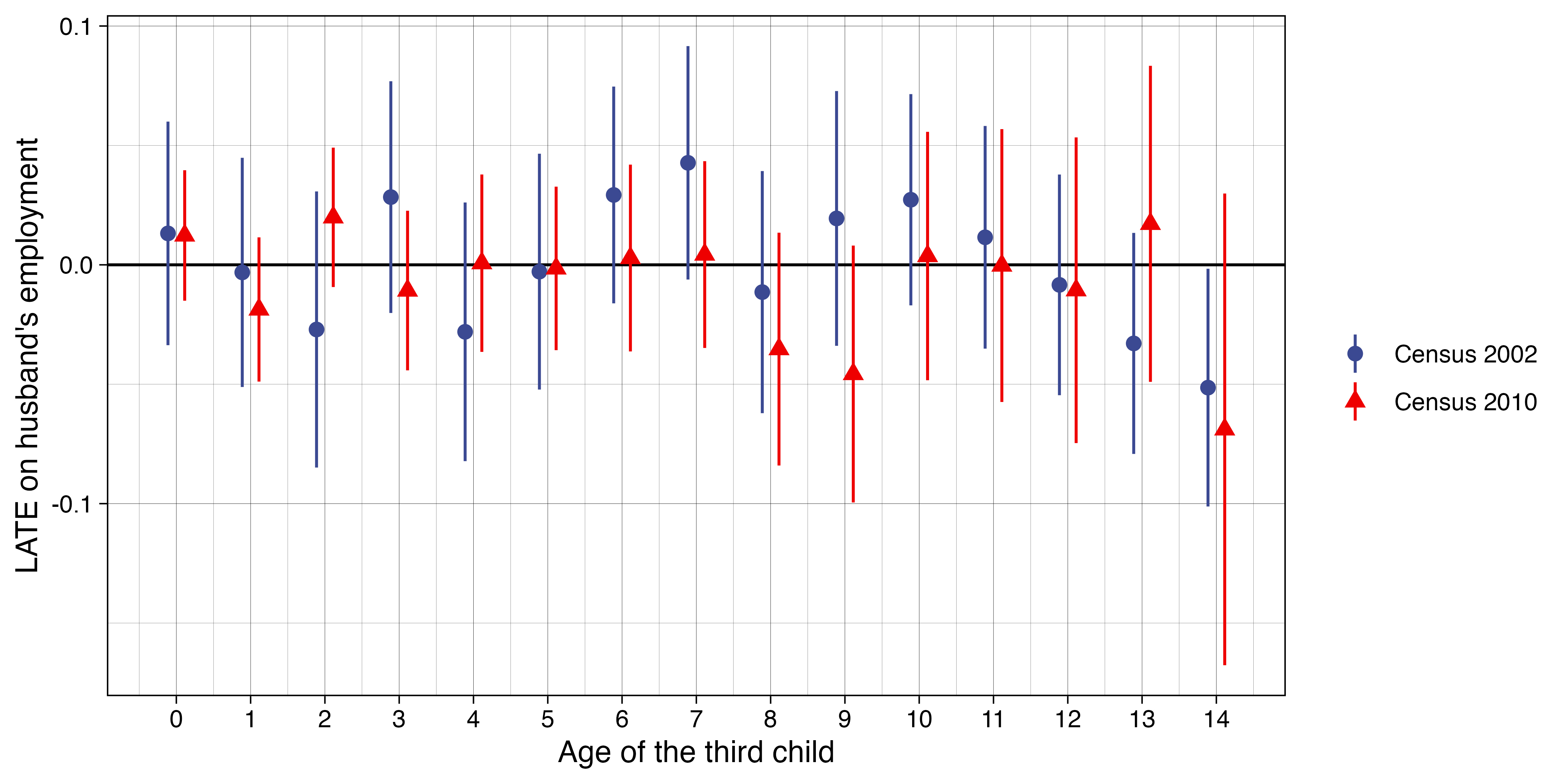}
        \caption{LATE of childbearing on husband's employment by age of the third child, census data.}
        \label{fig:late_on_empl_by_3age_census_fathers}
    \end{subfigure}
    \begin{minipage}{\textwidth}
        \caption{LATE of childbearing on mother's (a) and husband's (b) employment by age of the third child, census data.}
        \label{fig:late_on_empl_by_3age_census}
        \small{\textit{Note: Estimates are based on 2SLS models by sub-samples of the third child age with same sex and multiple births as instruments; covariates are mother's and husband's ages (categorical variable with 5-year age groups), children's age and children's sex; lines depict 95\% confidence interval constructed using heteroskedasticity consistent standard errors (HC1).}}
    \end{minipage}
\end{figure*}

Figure \ref{fig:late_on_empl_by_3age_census} examines the heterogeneity of the motherhood penalty by the age of the third child. Subfigures \subref{fig:late_on_empl_by_3age_census_mothers} and \subref{fig:late_on_empl_by_3age_census_fathers} shows this effect for mother and fathers respectively. Previous studies have demonstrated that the effect of motherhood should be strongest in the first years following childbirth, and then becomes weak. It is evident that, despite the relatively modest magnitude of the effect previously presented in Table \ref{tab:census2002_mothers_nocontrols} and \ref{tab:census2010_mothers_nocontrols}, women with young children encounter a dramatic decline in employment until approximately the age of 3. This phenomenon is well explained by the russian legislation that enables parental leave up to the child's age of 3. As displayed in the graph, this impact of childbirth on employment is fully extinguished after child reaches the age of 4. In other words, the average mother does not resign from her job after giving birth, but she takes legally established leave. However, it is worth noting that, apparently, not everyone takes advantage of this, since the effect is about 15\%, when child is 0-1 years old, while after that the effect is diminishing quickly. It is interesting to note that after the child starts going to school, the effect gradually moves again in the direction of decreasing female employment. The analysis of the two censuses yielded analogous results, indicating that after the child reaches the age of 12, a stable yet relatively modest effect of 5\% emerges. It is plausible that this is the long-term impact of motherhood on female employment. One potential explanation for the prolonged null effect of motherhood, which is evident when the child is aged between 3 and 11 years, may be that during this period, mothers are still predominantly attached to the labour market and are unwilling to make significant sacrifices, such as leaving their jobs. Secondly, this period is characterised by substantial expenses related to the child's upbringing, including kindergarten, primary education, and other related costs\footnote{However, the financial obligations associated with an older child must also be substantial, given the considerable costs associated with their education.}. The probability of this outcome being attributable to chance appears negligible, given the relatively high strength of the instruments and the presence of repeating patterns. Consequently, the present study recommends that this "pause in the motherhood penalty" be the subject of further detailed study.

In turn, Figure \ref{fig:late_on_empl_by_3age_census} (\subref{fig:late_on_empl_by_3age_census_fathers}) demonstrates the absence of heterogeneity of the effect depending on the age of the child for fathers, which once again supports Hypothesis 2 that fathers do not reduce their supply in the labor market. While the aforementioned analysis clearly demonstrates that the fatherhood does not result in a negative impact on employment outcomes, the question of the existence of a 'fatherhood premia' remains a subject of debate. It is evident that the vast majority of the male participants in the present study are employed, thereby hindering the possibility of investigating the hypothetical enhancement in labour supply. For this reason, Appendix \ref{app:census}, Table \ref{tab:census2010_secondjob} contains an additional analysis for women and men, which is similar to the analysis previously considered, but the presence of a second job is used as the dependent variable, and the sample is limited to working mothers and their partners\footnote{Unfortunately, there is no question about second job in the 2002 census, so only analysis with usage of 2010 data is presented.}. It is interesting to note that the 2SLS model with multiple birth as instruments supports the OLS finding, while the model with the same-sex instrument shows no effect. As previously stated, the multiple birth instrument is indicative of the fertility shock, thereby amplifying the impact of all factors. Notwithstanding, the robustness of this finding is apparent. Consequently, it can be tentatively proposed that in Russia, there is a negligible permanent fatherhood premia, evidenced by an increase in employment in the form of second jobs.

\subsection{Results with longitudinal data}\label{sec:results_rlms}

The present Section presents the results obtained from the analysis of panel survey data (RLMS, 2004-2024) using panel matching, which demonstrates the effect of childbearing on the economic behavior of women over a period of six years. As previously mentioned, it is crucial to interpret these results as supplementary, given the substantial set of strong assumptions they necessitate, which are challenging to verify. In this regard, the paper does not formally formulate causal hypotheses regarding the maternal penalty that could be tested on these data. Nevertheless, certain propositions were deduced that may facilitate a more profound interpretation of the results obtained from the census data. Firstly, the impact of childbirth on employment, monthly income, and average weekly working hours across several demographic groups is examined. The groups under comparison are defined by birth order, with first childless women being compared with women who have had their first child in the sample. Subsequently, a comparison is made between women with one child and those who experienced a birth of a second child. Thereafter, the comparison between women with two children and women who experienced a birth of a third child is shown. 

\begin{figure}[H] 
    \begin{minipage}{1\textwidth}
        \centering
        \includegraphics[width=1\textwidth]{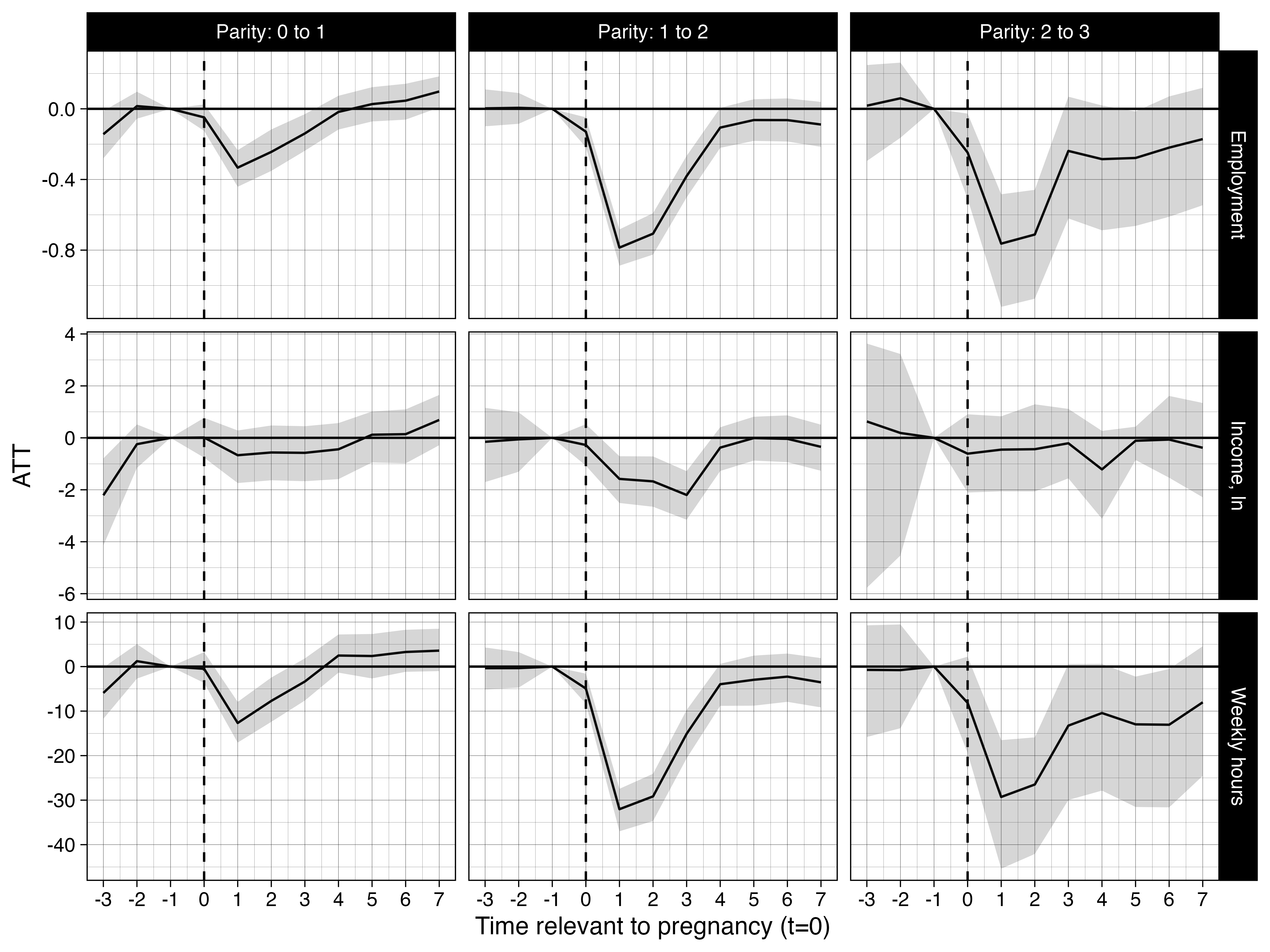}
        \caption{ATT of childbearing for mothers on different labor market outcomes by parity, RLMS 2004-2024.}
        \label{fig:rlms_all}
    \end{minipage}\hfill
    \small{\textit{Note: Estimates are based on panel matching with CBPS weighting; dashed area corresponds to empirical 95\% confidence interval constructed using 1000 iterations of blocked bootstrap; estimates for pre-treatment period are from placebo-tests, where $t=-1$ is taken as reference level.}}
\end{figure}

The primary findings are illustrated in Figure \ref{fig:rlms_all}, which comprises six graphs. The position of the figures is determined in such a way that the columns divide women by birth order, and the rows indicate the dependent variable. All of the graphs under consideration are constructed on the same model specification, with the exception of the dependent variable and birth order, which vary between them. As has been previously stated on numerous occasions, the fundamental assumption underpinning all the results presented in this section is the fulfilment of the PTA. This premise can be partially supported if the pre-natal effect is not significantly different from zero; otherwise, there is a high risk of selection bias, which hinders the interpretation of the results as causal. It is observed that the parallel trends are well-maintained for the second and third birth orders across all variables (see columns 2 and 3, periods from -3 to -1). However, a notable disparity emerges when the sample is divided into those who are childless and those who have given birth to their first child (see column 1). Here, a marked distinction is evident between the control and experimental groups in the initial period ($t = -3$). It is worth noting that subsequent periods show parallel trends, but it is writer's opinion that these estimates should be treated with a certain degree of caution. Indeed, the disparities between non-mothers and mothers are more pronounced than those observed between mothers with different numbers of children that explain the possible fail to control for selection bias for the first parity group. 

Returning to the results, the graph illustrates that the birth of a first, second or third child has a significant negative effect on women's employment until the child reaches the age of three (see the first row). This is most likely explained by the provision of parental leave as outlined above. It is also noteworthy that the census data yielded analogous results, yet panel matching is unable to provide a reliable indication of the termination of the "pause in the maternal penalty" previously identified (when there is a drop of employment at first 3 years after birth, than absence of effect, and finally small negative effect after child is 11 years old). A further intriguing question pertains to the magnitude of the effect. The first birth has an approximate 35\% negative effect on employment, while the second and third have a rate of approximately 80\%. These results are almost threefold those found with the census data. There are several potential explanations for such discrepancies. Firstly, the presented findings might be subject to bias due to the failure to meet the causal identification assumptions, resulting in inflated estimates. Secondly, it can be true that the LATE deviates from the ATT (while both are valid), a discrepancy that can be attributed to the distinct characteristics of the "compliers" population. Finally, it is possible to assume that the census the RLMS data demonstrate the impact of childbirth at different times: RLMS data examines primarily the period of 2010-2020, while censuses show 2002 and 2010 years. It is my belief that all three of these statements are true, however, it should be noted that there is more reason to trust the results obtained from the censuses.

Additionally, Figure \ref{fig:rlms_all} presents the findings on the effect of motherhood on income and the number of hours worked per week, which could not be studied using census data. It is unambiguous that the number of hours worked experiences a decline immediately following childbirth (see the third row), subsequently recovering and stabilizing. The most pronounced effects are observed for second and third births, which entirely replicate the results obtained with employment (that is the expected congruence). A more fruitful avenue of analysis would be to examine the dynamics of income (see the second row). Contrary to expectations, the birth of the first child does not have a significant effect on the mother's income, although a decrease in her activity in the labor market is observed. The most plausible explanation for this phenomenon is the role of government support and benefits, which serve to mitigate the impact of the decline in working hours. Conversely, when considering the dynamics of subsequent childbirths, a marked decline in income during the early years succeeding the birth of a child is evident, though this decline is subsequently stabilized. In turn, the birth of a third has no effect. A decline in income with second births generally corresponds with a significant decrease in labor market activity. However, in the case of third births, the effect is insignificant while there observed the same drop in employment, which is most likely explained by the small sample size that does not allow for the detection of this effect.

\begin{figure}[H] 
    \begin{minipage}{1\textwidth}
        \centering
        \includegraphics[width=0.8\textwidth]{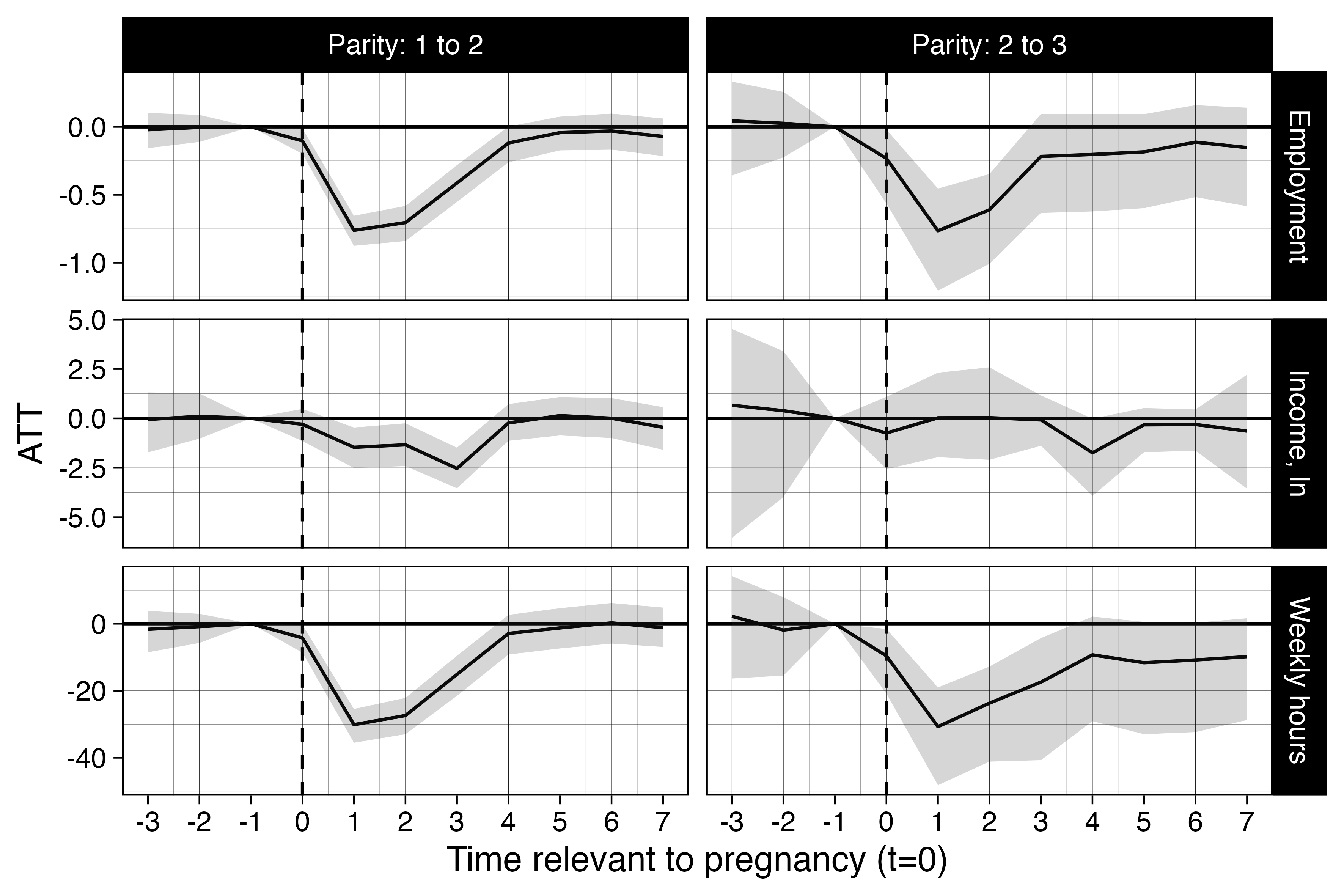}
        \caption{ATT of childbearing for mothers with partners on different labor market outcomes by parity, RLMS 2004-2024.}
        \label{fig:rlms_wife}
    \end{minipage}\hfill
    \small{\textit{Note: Estimates are based on panel matching with CBPS weighting; dashed area corresponds to empirical 95\% confidence interval constructed using 1000 iterations of blocked bootstrap; estimates for pre-treatment period are from placebo-tests, where $t=-1$ is taken as reference level.}}
\end{figure}

Figure \ref{fig:rlms_wife} explores the impact of childbearing by birth order on a sample of women in partnership. In this instance, the models encompass variables that reflect the male participant's involvement in the labor market, namely his employment status and earnings. However, the analysis for the first birth is not presented, due to the fact that the results obtained from the larger sample before did not allow for the identification of exhaustive evidence for PTA. All in all, the findings are evidently analogous to the outcomes observed for the complete sample discussed above. This is primarily due to the relatively low proportion of single mothers in the population. However, it is important to note that when the characteristics of the mother's partner are taken into account, it does not lead to any change to the results, which indicates the robustness of the findings with regard to different specifications.

\begin{figure}[H] 
    \begin{minipage}{1\textwidth}
        \centering
        \includegraphics[width=1\textwidth]{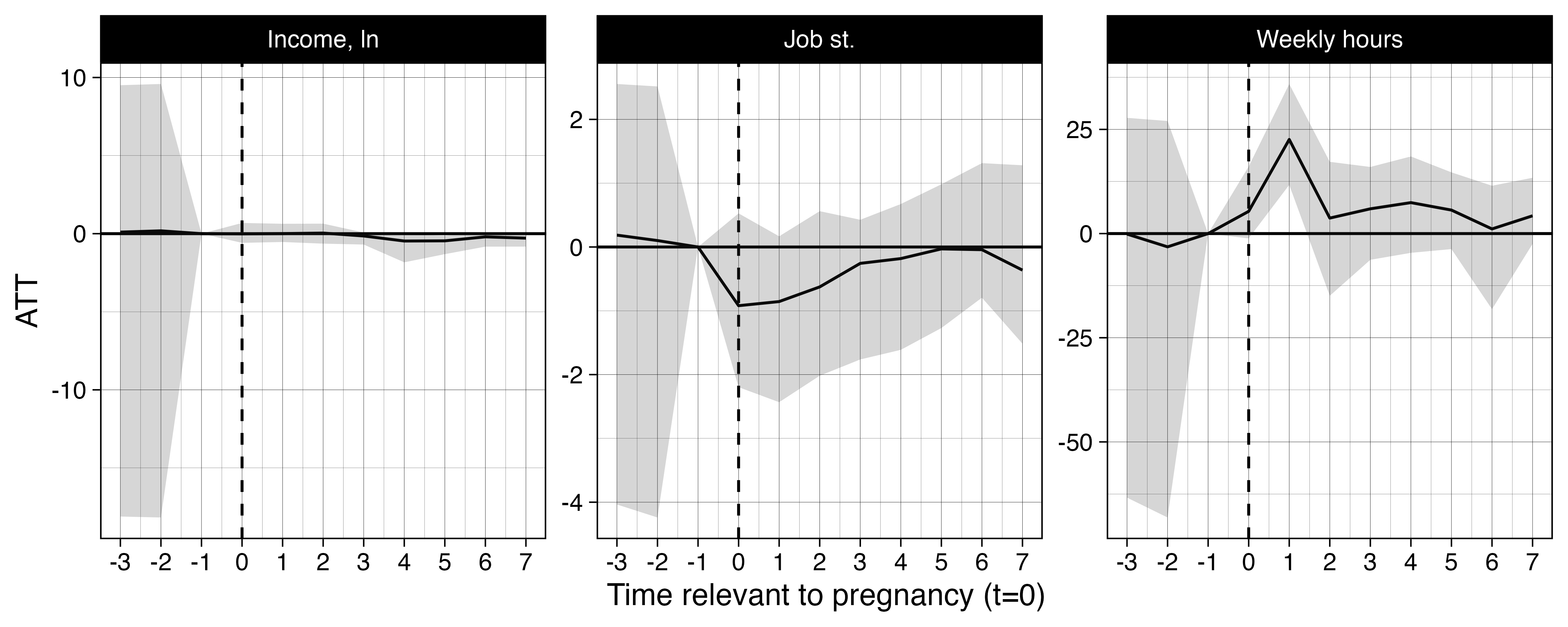}
        \caption{ATT of childbearing for employed mothers on different labor market outcomes, who experienced the second birth, RLMS 2004-2024.}
        \label{fig:rlms_empl}
    \end{minipage}\hfill
    \small{\textit{Note: Estimates are based on panel matching with CBPS weighting; dashed area corresponds to empirical 95\% confidence interval constructed using 1000 iterations of blocked bootstrap; estimates for pre-treatment period are from placebo-tests, where $t=-1$ is taken as reference level.}}
\end{figure}

Finally, Figure \ref{fig:rlms_empl} presents an analysis of the sample of working mothers, i.e. those who worked throughout the observation period (and therefore did not take maternity leave). There are not many of such women to conduct the analysis and select an adequate control group. As previously mentioned, the potential for selection bias led to the decision not to compare childless women with mothers who gave birth to their first child. Conversely, when comparing mothers with two children and those who gave birth to a third, the sample size of employed women is inadequate to detect any effect. Consequently, the graph displays a comparison of women with one and two children only. In contrast to the preceding analysis, it is seen that working mothers do not encounter a decline in their income that can be attributed to their decision not to take maternity leave. Furthermore, the graph demonstrates a marginal increase in the number of hours worked following childbirth. Finally, the findings reveal that, contrary to the assumptions put forward in the theoretical section, women's job satisfaction remains at the same level. This may be explained by the fact that women continued to be employed after having children because their high levels of job satisfaction. In other words, it is job satisfaction that might be a factor to continue to work after childbearing. 

\section{Discussion}\label{sec:discussion}

The study's initial inquiry concerned the extent to which the findings of studies conducted in other countries can be applied to Russia. In other words, does Russia have specific features in the parental penalty? A review of earlier research conducted on Russian data indicates that the financial penalty for motherhood in Russia is only marginally lower than in developed countries. However, the findings of this study suggest an alternative response to this question: in Russia, the financial penalties associated with motherhood, at least in terms of employment, vary significantly compared to those observed in developed countries and in those few developing countries for which studies are available. Consequently, the impact of childbirth, as evidenced by studies employing analogous methodologies, is estimated to be approximately 10\% \citep{meta1,angrist_orig,meta2}, with this effect persisting until the child attains adulthood. In contrast, for the Russian context, it is argued that the effect is no more than 5\%, with the predominant reduction in employment occurring in the first three years following the birth of the child. Furthermore, the investigation revealed the absence of a significant fatherhood premium in terms of employment in Russia, although a modest increase in labor supply was observed in instances of secondary employment. The veracity of these results is confirmed through the application of an alternative set of data sources and methodologies.

Moreover, the study demonstrates that, in contrast to the findings of preceding research, in Russia the greater penalty for motherhood is more likely to be associated with women employed in low-skilled labor sectors, as evidenced by a comparison of the effect of motherhood with educational attainment. However, empirical studies conducted in developed countries have yielded contrasting results, findings that are corroborated by a comprehensive theory of human capital. I propose explanations that are based on unique russian institutional context. Namely, women who have received a higher education are more likely to be employed within the public sector, which is a "family-friendly" employer. Secondly, due its strict russian labor law, employers are likely to fire women with children, but mothers who have received tertiary education are more likely to be able to defend their rights before their employers.

Finally, it is important to reemphasize the limitations of this study once again. The primary source of data is Russian census data; however, a number of authors have raised valid concerns regarding its reliability. This may result in a lower external validity of the present study. The second limitation that merits mention is the effect found using instrumental variables, which is a local average treatment effect, as opposed to the average treatment effect. The study examines the impact of childbearing on individuals whose treatment status (decision to have one more child) can be influenced by the instrument (siblings sex composition or the occurrence of multiple births). Therefore, the study focuses on a subpopulation of individuals who are known as "compliers". As outlined above, the proportion of such compliers is not large. However, in line with the contributions of preceding researchers, there is no compelling evidence to suggest that the "compliers" exhibit significant disparities when compared to the broader population, thereby rendering the findings pertinent to the general case.

The subsequent constraint of the present study is its design. On the one hand, instrumental variables enable the formulation of causal and sufficiently reliable inferences, thus facilitating a transition from correlations. Conversely, this approach entails a number of costs, primarily the necessity for a specific sample. In the theoretical section, the general effect of having a child was considered, which includes the first, second, and subsequent births. However, in the main analysis, comparison is drawn between women with two and three children. Given the current fertility trends in Russia, the number of women with three children is decreasing, which calls into question the interpretation of the results as a general penalty for motherhood, and suggests that the estimated effect may be associated specifically with the third child. However, an additional analysis of the RLMS data reveals that the impact of motherhood when comparing the second and third birth is almost identical. This finding enables some extrapolation of the effect to at least second births, contingent on certain assumptions. In the context of the first childbirth, the situation is considerably more complex, because groups of women with one and three children should differ dramatically. However, under the assumption that the motherhood penalty is at least somewhat linear in relation to the number of children, the identified effect can be regarded as its most precise estimation.

\section{Conclusion}\label{sec:conclusion}

The present study aimed to improve upon the existing correlational literature on the parenthood penalty in Russia. An instrumental variables approach based on sibling sex composition and multiple births was employed alongside difference-in-differences designs to analyze rich census and longitudinal datasets. To the best of the authors' knowledge, this is the first study to provide causal estimates of the effect of fertility decisions on subsequent labor market outcomes for mothers and fathers in contemporary Russia. The study's primary finding is that, in contrast to the approximately 10 percent long-term motherhood penalty observed in developed countries, the causal impact of childbearing on women's employment in Russia is most significant in the first year after birth, reducing employment by around 15 percent. This penalty then rapidly declines to a modest 3 percent once children reach school age. The analysis indicates an absence of a systematic fatherhood penalty in terms of employment, although a modest increase in labor supply is observed.

It is important to note that the 'motherhood penalty' varies across different occupational groups. Evidence suggests that women in low-skilled roles experience a substantially greater decline in employment, while those with a tertiary education encounter a comparatively minor impact. This phenomenon can be attributed to their comparatively higher tendency to occupy secure, family-friendly positions within the public sector, along with their enhanced capacity to uphold employee rights granted by strict protective laws in Russia that are characterized by low enforcement. These findings contradict the predictions of standard human capital models, emphasizing the importance of Russia's unique institutional background. This includes the legacy of Soviet-era labor force norms, the structure of public sector employment, stringent labor law protections and recent pronatalist policy interventions.

\bibliography{references}

\newpage
\begin{appendices}
\section*{Appendix}
\section{Additional results with census data}\label{app:census}

\begin{table}[!h]

\caption{\label{tab:balance_census_ss}Descriptive statistic by instruments' values}
\begin{tabular}[t]{lcccc}
\toprule
\multicolumn{1}{c}{ } & \multicolumn{1}{c}{Whole sample} & \multicolumn{3}{c}{Same sex} \\
\cmidrule(l{3pt}r{3pt}){2-2} \cmidrule(l{3pt}r{3pt}){3-5}
Variable & Mean (sd) & Z = 1 & Z = 0 & Diff. (t-stat)\\
\midrule[1.5pt] \multicolumn{5}{c}{\textbf{Census 2002}} \\ \midrule[1.5pt]
\midrule
Mother's age & 35.19
(5.92) & 35.19
(5.94) & 35.19
(5.91) & 0.005
(0.33)\\
\addlinespace[0.5em]
Age at 1st birth & 22.66
(3.91) & 22.67
(3.92) & 22.65
(3.9) & 0.0265
(2.65)\\
\addlinespace[0.5em]
Graduated & 0.22
(0.41) & 0.22
(0.41) & 0.22
(0.41) & -3e-04
(-0.28)\\
\addlinespace[0.5em]
Married & 0.83
(0.38) & 0.82
(0.38) & 0.83
(0.38) & -0.0037
(-3.77)\\
\addlinespace[0.5em]
Husband's age & 37.44
(6.07) & 37.45
(6.09) & 37.44
(6.06) & 0.0132
(0.77)\\
\addlinespace[0.5em]
Number of children & 2.19
(0.49) & 2.21
(0.51) & 2.17
(0.47) & 0.0353
(28.01)\\
\addlinespace[0.5em]
More than 2 children & 0.15
(0.36) & 0.17
(0.37) & 0.14
(0.35) & 0.0286
(31.16)\\
\addlinespace[0.5em]
1st child's sex & 0.52
(0.5) & 0.53
(0.5) & 0.51
(0.5) & 0.0212
(16.6)\\
\addlinespace[0.5em]
2nd child's sex & 0.51
(0.5) & 0.53
(0.5) & 0.49
(0.5) & 0.0349
(27.31)\\
\addlinespace[0.5em]
1sr child's age & 12.53
(4.14) & 12.52
(4.16) & 12.54
(4.13) & -0.0215
(-2.03)\\
\addlinespace[0.5em]
2nd child's age & 8.35
(4.82) & 8.38
(4.83) & 8.33
(4.81) & 0.049
(3.97)\\
\midrule[1.5pt] \multicolumn{5}{c}{\textbf{Census 2010}} \\ \midrule[1.5pt]
\addlinespace[0.5em]
Mother's age & 33.72
(5.51) & 33.72
(5.53) & 33.72
(5.49) & 0.0038
(0.25)\\
\addlinespace[0.5em]
Age at 1st birth & 22.59
(3.84) & 22.61
(3.86) & 22.57
(3.82) & 0.0436
(4.09)\\
\addlinespace[0.5em]
Graduated & 0.33
(0.47) & 0.34
(0.47) & 0.33
(0.47) & 3e-04
(0.23)\\
\addlinespace[0.5em]
Married & 0.82
(0.38) & 0.82
(0.38) & 0.83
(0.38) & -0.0049
(-4.61)\\
\addlinespace[0.5em]
Husband's age & 36.54
(6.34) & 36.53
(6.34) & 36.55
(6.35) & -0.0249
(-1.28)\\
\addlinespace[0.5em]
Number of children & 2.2
(0.5) & 2.22
(0.52) & 2.19
(0.48) & 0.0337
(24.27)\\
\addlinespace[0.5em]
More than 2 children & 0.17
(0.37) & 0.18
(0.38) & 0.15
(0.36) & 0.0272
(26.32)\\
\addlinespace[0.5em]
1st child's sex & 0.52
(0.5) & 0.53
(0.5) & 0.51
(0.5) & 0.0176
(12.71)\\
\addlinespace[0.5em]
2nd child's sex & 0.51
(0.5) & 0.53
(0.5) & 0.49
(0.5) & 0.041
(29.51)\\
\addlinespace[0.5em]
1sr child's age & 11.13
(4.53) & 11.11
(4.55) & 11.15
(4.52) & -0.0398
(-3.16)\\
\addlinespace[0.5em]
2nd child's age & 5.98
(4.51) & 6
(4.53) & 5.96
(4.5) & 0.0394
(3.14)\\
\bottomrule
\end{tabular}
\end{table}

\begin{table}[!h]

\caption{\label{tab:balance_census_mb}Descriptive statistic by instruments' values}
\begin{tabular}[t]{lcccc}
\toprule
\multicolumn{1}{c}{ } & \multicolumn{1}{c}{Whole sample} & \multicolumn{3}{c}{Multiple births} \\
\cmidrule(l{3pt}r{3pt}){2-2} \cmidrule(l{3pt}r{3pt}){3-5}
Variable & Mean (sd) & Z = 1 & Z = 0 & Diff. (t-stat)\\
\midrule[1.5pt] \multicolumn{5}{c}{\textbf{Census 2002}} \\ \midrule[1.5pt]
\midrule
Mother's age & 35.19
(5.92) & 35.81
(5.95) & 35.18
(5.92) & 0.632
(7.33)\\
\addlinespace[0.5em]
Age at 1st birth & 22.66
(3.91) & 23.02
(4.06) & 22.66
(3.91) & 0.3601
(6.12)\\
\addlinespace[0.5em]
Graduated & 0.22
(0.41) & 0.2
(0.4) & 0.22
(0.41) & -0.013
(-2.22)\\
\addlinespace[0.5em]
Married & 0.83
(0.38) & 0.82
(0.38) & 0.83
(0.38) & -0.0064
(-1.15)\\
\addlinespace[0.5em]
Husband's age & 37.44
(6.07) & 37.9
(6.23) & 37.44
(6.07) & 0.468
(4.7)\\
\addlinespace[0.5em]
Number of children & 2.19
(0.49) & 3.14
(0.41) & 2.18
(0.48) & 0.957
(162.39)\\
\addlinespace[0.5em]
More than 2 children & 0.15
(0.36) & 1
(0) & 0.14
(0.35) & 0.8571
(1905.67)\\
\addlinespace[0.5em]
1st child's sex & 0.52
(0.5) & 0.52
(0.5) & 0.52
(0.5) & 0.002
(0.28)\\
\addlinespace[0.5em]
2nd child's sex & 0.51
(0.5) & 0.49
(0.5) & 0.51
(0.5) & -0.0193
(-2.67)\\
\addlinespace[0.5em]
1sr child's age & 12.53
(4.14) & 12.8
(4.05) & 12.53
(4.15) & 0.2719
(4.63)\\
\addlinespace[0.5em]
2nd child's age & 8.35
(4.82) & 8.52
(4.79) & 8.34
(4.81) & 0.1897
(2.74)\\
\midrule[1.5pt] \multicolumn{5}{c}{\textbf{Census 2010}} \\ \midrule[1.5pt]
\addlinespace[0.5em]
Mother's age & 33.72
(5.51) & 34.12
(5.81) & 33.72
(5.51) & 0.3991
(4.55)\\
\addlinespace[0.5em]
Age at 1st birth & 22.59
(3.84) & 23.1
(4.25) & 22.58
(3.83) & 0.5125
(7.99)\\
\addlinespace[0.5em]
Graduated & 0.33
(0.47) & 0.35
(0.48) & 0.33
(0.47) & 0.019
(2.62)\\
\addlinespace[0.5em]
Married & 0.82
(0.38) & 0.83
(0.37) & 0.82
(0.38) & 0.0114
(2.03)\\
\addlinespace[0.5em]
Husband's age & 36.54
(6.34) & 36.95
(6.59) & 36.53
(6.34) & 0.4164
(3.82)\\
\addlinespace[0.5em]
Number of children & 2.2
(0.5) & 3.12
(0.38) & 2.19
(0.49) & 0.9294
(160.61)\\
\addlinespace[0.5em]
More than 2 children & 0.17
(0.37) & 1
(0) & 0.16
(0.36) & 0.8442
(1664.98)\\
\addlinespace[0.5em]
1st child's sex & 0.52
(0.5) & 0.52
(0.5) & 0.52
(0.5) & -3e-04
(-0.03)\\
\addlinespace[0.5em]
2nd child's sex & 0.51
(0.5) & 0.5
(0.5) & 0.51
(0.5) & -0.013
(-1.73)\\
\addlinespace[0.5em]
1sr child's age & 11.13
(4.53) & 11.02
(4.65) & 11.13
(4.53) & -0.1134
(-1.61)\\
\addlinespace[0.5em]
2nd child's age & 5.98
(4.51) & 5.7
(4.43) & 5.96
(4.5) & -0.2654
(-3.97)\\
\bottomrule
\end{tabular}
\end{table}

\begin{sidewaystable}
    \caption{OLS and 2SLS estimates of the effect of childbearing on mother's and husband's second job, 2010 census.}
    
\begin{threeparttable}
\begin{tabular}{l D{)}{)}{9)3} D{)}{)}{10)6} D{)}{)}{8)0} D{)}{)}{9)0} D{)}{)}{9)3} D{)}{)}{10)7} D{)}{)}{9)3} D{)}{)}{9)3}}
\toprule
 & \multicolumn{4}{c}{Mothers} & \multicolumn{4}{c}{Husbands} \\
\cmidrule(lr){2-5} \cmidrule(lr){6-9}
 & \multicolumn{1}{c}{OLS} & \multicolumn{1}{c}{2SLS} & \multicolumn{1}{c}{2SLS} & \multicolumn{1}{c}{2SLS} & \multicolumn{1}{c}{OLS} & \multicolumn{1}{c}{2SLS} & \multicolumn{1}{c}{2SLS} & \multicolumn{1}{c}{2SLS} \\
\midrule
Intercept            & 0.06 \; (0.04)                 & 0.06 \;  (0.04)                           & 0.06 \; (0.04)                              & 0.06 \; (0.04)                              & 0.07 \; (0.02)^{***}           & 0.08 \;  (0.02)^{***}                      & 0.07 \; (0.02)^{***}                        & 0.07 \; (0.02)^{***}                        \\
3 children & 0.01 \; (0.00)^{***}           & 0.03 \;  (0.03)                           & 0.03 \; (0.03)                              & 0.01 \; (0.01)                              & 0.01 \; (0.00)^{***}           & 0.03 \;  (0.02)                            & 0.03 \; (0.02)                              & 0.01 \; (0.00)^{***}                        \\
1st child sex        & 0.00 \; (0.00)                 & 0.00 \;  (0.00)                           & 0.00 \; (0.00)                              & 0.00 \; (0.00)                              & -0.00 \; (0.00)                & -0.00 \;  (0.00)                           & -0.00 \; (0.00)                             & -0.00 \; (0.00)                             \\
2nd child sex        & -0.00 \; (0.00)                & -0.00 \;  (0.00)                          &                                             & -0.00 \; (0.00)                             & -0.00 \; (0.00)                & -0.00 \;  (0.00)                           &                                             & -0.00 \; (0.00)                             \\
\midrule
IV                   & \multicolumn{1}{c}{\small{--}} & \multicolumn{1}{c}{\small{Same sex}}      & \multicolumn{1}{c}{\small{Both boys/girls}} & \multicolumn{1}{c}{\small{Multiple births}} & \multicolumn{1}{c}{\small{--}} & \multicolumn{1}{c}{\small{Same sex}}       & \multicolumn{1}{c}{\small{Both boys/girls}} & \multicolumn{1}{c}{\small{Multiple births}} \\
Weak Instr.          & \multicolumn{1}{c}{\small{--}} & \multicolumn{1}{c}{\small{354.27 ***}}    & \multicolumn{1}{c}{\small{177.8 ***}}       & \multicolumn{1}{c}{\small{18017.21 ***}}    & \multicolumn{1}{c}{\small{--}} & \multicolumn{1}{c}{\small{569.34 ***}}     & \multicolumn{1}{c}{\small{286.45 ***}}      & \multicolumn{1}{c}{\small{22977.7 ***}}     \\
AR                   & \multicolumn{1}{c}{\small{--}} & \multicolumn{1}{c}{\small{(-0.03) - 0.1}} & \multicolumn{1}{c}{\small{--}}              & \multicolumn{1}{c}{\small{0 - 0.02}}        & \multicolumn{1}{c}{\small{--}} & \multicolumn{1}{c}{\small{(-0.01) - 0.08}} & \multicolumn{1}{c}{\small{--}}              & \multicolumn{1}{c}{\small{0.01 - 0.02}}     \\
Wu-Hausman           & \multicolumn{1}{c}{\small{--}} & \multicolumn{1}{c}{\small{0.72  }}        & \multicolumn{1}{c}{\small{0.64  }}          & \multicolumn{1}{c}{\small{0.11  }}          & \multicolumn{1}{c}{\small{--}} & \multicolumn{1}{c}{\small{1.27  }}         & \multicolumn{1}{c}{\small{1.1  }}           & \multicolumn{1}{c}{\small{4.85 *}}          \\
Sargan               & \multicolumn{1}{c}{\small{--}} & \multicolumn{1}{c}{\small{--}}            & \multicolumn{1}{c}{\small{0.56  }}          & \multicolumn{1}{c}{\small{--}}              & \multicolumn{1}{c}{\small{--}} & \multicolumn{1}{c}{\small{--}}             & \multicolumn{1}{c}{\small{1.06  }}          & \multicolumn{1}{c}{\small{--}}              \\
Num. obs.            & 232107                         & 232107                                    & 232107                                      & 231487                                      & 347970                         & 347970                                     & 347970                                      & 346956                                      \\
\bottomrule
\end{tabular}
\begin{tablenotes}[flushleft]
\small{\item Note: $^{***}p<0.001$; $^{**}p<0.01$; $^{*}p<0.05$. The analysis is conducted on sub-samples of working mothers or husbands; as control variables in all models are included mother's and husband's ages (categorical variables with 5-year age groups), children's age and rural dummy as well as region FE; heteroskedasticity consistent standard errors (HC1) are in parentheses; test on weak instruments shows robust F-statistic from the first-stage; AR - Anderson-Rubin 95-CI.}
\end{tablenotes}
\end{threeparttable}
 
    \label{tab:census2010_secondjob}
\end{sidewaystable}

\newpage
\section{Additional results with RLMS data}\label{app:rlms}

\addtolength{\tabcolsep}{+3pt}  

\begin{ThreePartTable}
\begin{TableNotes}
\item \textit{Note: } 
\item N ch. - number of children; Empl. - employment; sat. - satisfaction; Income is in constant 2022 rubles. Means are weighted by surbey weights.
\end{TableNotes}
\begin{longtable}[t]{ccccccccc}
\caption{\label{tab:rlms_desc}Descriptive statistic on women in RLMS data, means.}\\
\toprule
Year & N & Age & N ch. & Empl. & Income & Hours worked & Job sat. & Graduated\\
\midrule
2004 & 4535 & 39.76 & 1.35 & 0.59 & 17178.61 & 24.42 & 2.86 & 0.22\\
\addlinespace[0.5em]
2005 & 4294 & 39.35 & 1.32 & 0.60 & 18651.85 & 24.61 & 2.84 & 0.22\\
\addlinespace[0.5em]
2006 & 5404 & 39.10 & 1.28 & 0.62 & 23116.17 & 25.23 & 2.62 & 0.24\\
\addlinespace[0.5em]
2007 & 5279 & 39.33 & 1.28 & 0.62 & 24699.18 & 25.91 & 2.54 & 0.25\\
\addlinespace[0.5em]
2008 & 5061 & 39.30 & 1.26 & 0.63 & 29002.85 & 25.84 & 2.48 & 0.25\\
\addlinespace[0.5em]
2009 & 5217 & 39.63 & 1.27 & 0.63 & 30675.26 & 25.85 & 2.42 & 0.26\\
\addlinespace[0.5em]
2010 & 6947 & 39.99 & 1.31 & 0.60 & 27799.27 & 24.92 & 2.39 & 0.25\\
\addlinespace[0.5em]
2011 & 8453 & 40.87 & 1.32 & 0.60 & 29383.26 & 24.42 & 2.38 & 0.28\\
\addlinespace[0.5em]
2012 & 9159 & 41.46 & 1.34 & 0.60 & 35379.22 & 24.54 & 2.29 & 0.30\\
\addlinespace[0.5em]
2013 & 7727 & 41.61 & 1.35 & 0.60 & 34986.27 & 24.52 & 2.26 & 0.30\\
\addlinespace[0.5em]
2014 & 7051 & 41.54 & 1.33 & 0.60 & 34616.49 & 24.79 & 2.27 & 0.30\\
\addlinespace[0.5em]
2015 & 6522 & 41.62 & 1.32 & 0.59 & 31602.23 & 23.77 & 2.31 & 0.31\\
\addlinespace[0.5em]
2016 & 6812 & 41.42 & 1.33 & 0.59 & 31372.82 & 23.60 & 2.26 & 0.32\\
\addlinespace[0.5em]
2017 & 6668 & 41.48 & 1.31 & 0.59 & 35021.39 & 23.80 & 2.26 & 0.32\\
\addlinespace[0.5em]
2018 & 6305 & 41.53 & 1.30 & 0.59 & 32359.50 & 23.99 & 2.23 & 0.32\\
\addlinespace[0.5em]
2019 & 6233 & 41.59 & 1.30 & 0.60 & 33432.66 & 24.21 & 2.23 & 0.32\\
\addlinespace[0.5em]
2020 & 6217 & 41.44 & 1.27 & 0.59 & 34158.76 & 23.60 & 2.18 & 0.34\\
\addlinespace[0.5em]
2021 & 6111 & 42.71 & 1.35 & 0.60 & 35086.03 & 24.58 & 2.17 & 0.34\\
\addlinespace[0.5em]
2022 & 5838 & 42.86 & 1.37 & 0.62 & 34244.17 & 24.69 & 2.11 & 0.34\\
\addlinespace[0.5em]
2023 & 5777 & 43.02 & 1.36 & 0.63 & 36832.62 & 25.17 & 2.08 & 0.35\\
\addlinespace[0.5em]
2024 & 391 & 43.39 & 1.33 & 0.66 & 44782.31 & 26.17 & 2.03 & 0.44\\
\bottomrule
\insertTableNotes
\end{longtable}
\end{ThreePartTable}

\addtolength{\tabcolsep}{-3pt}

\end{appendices}
\end{document}